\documentclass[twocolumn]{aastex63}

\usepackage{graphicx}
\usepackage{dcolumn}
\usepackage{bm}
\usepackage[varvw]{newtxmath}
\usepackage{ulem}
\usepackage{afterpage}

\begin{document}

\shorttitle{The Galactic Center Reference Frame}
\shortauthors{Darling, Paine, Reid, Menten, Sakai, \& Ghez}

\title{An Updated Reference Frame for the Galactic Inner Parsec}

\author{Jeremy Darling}
 \affiliation{Center for Astrophysics and Space Astronomy,
Department of Astrophysical and Planetary Sciences\\
University of Colorado, 389 UCB,
Boulder, CO 80309-0389, USA}
 \email{jeremy.darling@colorado.edu}

\author{Jennie Paine}
 \affiliation{Center for Astrophysics and Space Astronomy,
Department of Astrophysical and Planetary Sciences\\
University of Colorado, 389 UCB,
Boulder, CO 80309-0389, USA}
 
 \author{Mark J. Reid}
 \affiliation{Center for Astrophysics,
   Harvard \& Smithsonian,
   60 Garden Street,
   Cambridge, MA 02138, USA}

 \author{Karl M. Menten}
 \affiliation{Max-Planck-Institut f\"{u}r Radioastronomie, Auf dem H\"{u}gel 69, D-53121 Bonn, Germany}

 \author{Shoko Sakai}
 \affiliation{UCLA Department of Physics and Astronomy, Los Angeles, CA 90095-1547, USA}

 \author{Andrea Ghez}
  \affiliation{UCLA Department of Physics and Astronomy, Los Angeles, CA 90095-1547, USA}

\begin{abstract}
  Infrared observations of stellar orbits about Sgr~A* probe the mass distribution in the inner parsec of the Galaxy and provide definitive evidence for the existence of a massive black hole.
  However, the infrared astrometry is relative and is tied to the radio emission from Sgr A* using stellar SiO masers that coincide with infrared-bright stars.
  To support and improve this two-step astrometry,
  we present new astrometric observations of 15 stellar SiO masers within 2 pc of Sgr~A*.  Combined with legacy observations spanning 25.8 years,
  we re-analyze the relative offsets of these masers from Sgr~A* and measure positions and proper motions
  that are significantly improved compared to the previously published reference frame.
  Maser positions are corrected for epoch-specific differential aberration, precession, nutation, and solar gravitational deflection.
  Omitting the supergiant IRS~7, the mean position uncertainties are 0.46~mas and 0.84~mas
  in RA and Dec., and the mean proper motion uncertainties are 0.07~mas~yr$^{-1}$ and 0.12~mas~yr$^{-1}$, respectively.
  At a distance of 8.2~kpc, these correspond to position uncertainties of 3.7 AU and 6.9 AU and proper motion
  uncertainties of 2.7~km~s$^{-1}$ and 4.6~km~s$^{-1}$.
  The reference frame stability, the uncertainty in the variance-weighted mean proper motion of the maser ensemble,
 is 8~$\mu$as~yr$^{-1}$ (0.30~km~s$^{-1}$) in RA and 11~$\mu$as~yr$^{-1}$ (0.44~km~s$^{-1}$) in Dec., which 
 represents a 2.3-fold improvement over previous work and a new benchmark for the maser-based reference frame.
 \end{abstract}

\section{\label{sec:intro}Introduction}

Infrared observations of stellar orbits in the vicinity of Sgr~A* spanning nearly three decades have demonstrated the presence of a massive black hole in the
Galactic Center \citep[e.g.,][]{ghez2008,genzel2010}.
These observations can also probe the mass distribution in the inner parsec, including that of the dark matter and other unseen material \citep{lacroix2018,nampalliwar2021,heissel2022,yuan2022}.
The infrared astrometry 
has historically relied on a radio-based astrometric reference frame that ties IR-bright stars to the location of Sgr~A* via simultaneous observation of SiO maser-emitting stars
and the Sgr~A* 43~GHz radio continuum \citep[e.g.,][]{menten1997,reid2003,reid2007,yelda2010,plewa2015,sakai2019}.
The predicted positions of these jointly-detected stars degrade over time, and the maser-based reference frame must therefore be regularly monitored and updated.  It has
now been 16 years since the last published maser observations used for the
Galactic Center reference frame \citep[][but see \citealt{sakai2019}]{reid2007}.

Here we present an updated radio reference frame for the Galactic Center that incorporates new and legacy
Karl G. Jansky Very Large Array (VLA\footnote{The National Radio Astronomy Observatory is a facility of the National Science Foundation operated under cooperative agreement by Associated Universities, Inc.}) data (Section \ref{sec:data}).  We employ new astrometric methods (Section \ref{sec:methods})
to obtain unprecedented position and proper motion measurements and reference frame stability (Section \ref{sec:results}). 
We examine the error budgets, systematic effects, and possible intrinsic scatter in the astrometry (Section \ref{sec:discussion}),
examine trends in the 3D stellar velocities (Section \ref{sec:analysis}), and discuss future
work (Section \ref{sec:conclusions}). The Appendices discuss time-dependent differential astrometric corrections, 
provide the complete maser time series, examine alternative proper motion fitting methods, and assess
the possibility of under-estimated astrometric uncertainties.

Calculations that convert angular offsets to projected physical distances or proper motions to transverse velocities assume a distance to Sgr~A* of 8.2~kpc,
which is consistent with most recent distance measurements \citep[e.g.,][]{do2019,reid2019,gravity2021,leung2023}.

\section{\label{sec:data} Data}

Table \ref{tab:data} lists the epochs, observing programs, observed SiO transitions, and beam properties of the legacy
and new data sources used to derive astrometric solutions for the stellar SiO masers near Sgr~A*.  There are additional masers
in the field of view, such as those detected by \citet{li2010} and \citet{paine2022}, as well as additional maser transitions
that are not included in this study because they do not have legacy astrometry \citep{reid2003,reid2007}.

\begin{deluxetable}{clccc}
  \tablecaption{\label{tab:data} SiO Maser Data and Observations}
  \tablehead{ \colhead{Mean Epoch} & \colhead{Program} & \colhead{$v$\tablenotemark{a}} & \colhead{Beam} & \colhead{Ref.} \\
    & & & (mas)}
  \startdata
1996.413  & VLBA  BM060  & 1 & $1.2\times0.9$  & R03 \\                
1998.410  & VLA AM592 & 1 & $70\times30$ & R03 \\          
2000.850  & VLA AR451 & 1 & $80\times40$ & R03 \\
2006.200  & VLA AR588 & 1 &  $86 \times 33$ & R07\\
2008.860  & VLA AR678 & 1 & $82\times35$ & R22 \\
2011.470  & VLA 11A-101 & 1 & $97\times42$ & R22 \\
2014.249 & VLA 14A-168 & 1,2 & $66\times30$ & R22 \\
2020.988 & VLA 19A-310 & 1,2 &$93\times36$ & P22,D22 \\
2022.227 & VLA 22A-328 & 1,2 & $82\times39$ & D22 \\
  \enddata
  \tablerefs{R03 = \citet{reid2003}; R07 = \citet{reid2007};
    R22 = Reid (2022, private communication); P22 = \citet{paine2022}; D22 = this work.}
  \tablenotetext{a}{Vibration quantum number (all transitions are $J=1-0$).}
\end{deluxetable}

\begin{figure*}[t]
\begin{centering}
  \includegraphics[scale=0.48,trim= 20 20 20 20,clip=false]{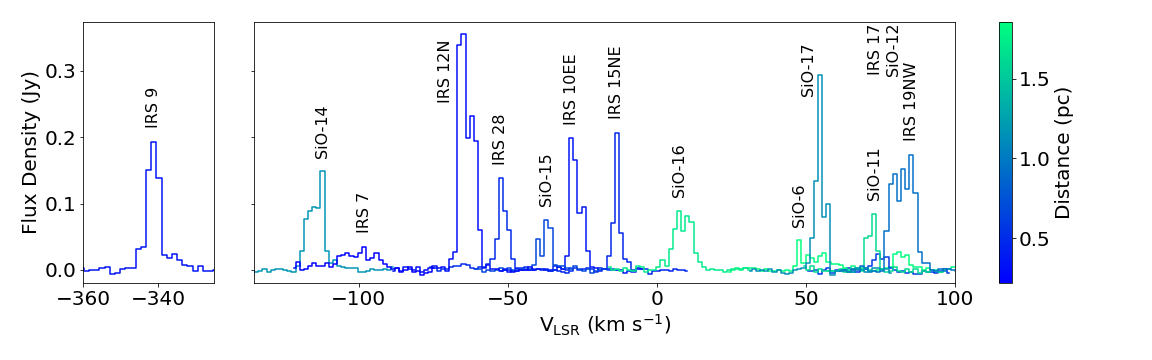}
  \caption{Composite spatially-integrated SiO $v=1$, $J=1-0$ maser spectra from the 2022.227 epoch image cubes.  The colors indicate the projected distance from Sgr~A*, assuming a Galactic Center distance
    of 8.2 kpc.}
\label{fig:spectra}
\end{centering}
\end{figure*}

\subsection{\label{subsec:legacy} Legacy Data}

We employ the VLA and VLBA astrometric measurements of 15 SiO masers presented in \citet{reid2003} and \citet{reid2007}.  These span 1996--2006.
In addition, we use the measurements obtained from VLA programs in 2008, 2011, and 2014 by Reid (2022, private communication).  
We did not use the 1995 VLA data presented in \citet{menten1997} because the uncertainties in the measured coordinates are an order of
magnitude larger than subsequent epochs due to larger synthesized beams.

\subsection{\label{subsec:obs} New Observations and Data Reduction}

New VLA observations were conducted in programs 19A-310 (27 Dec 2020 or 2020.988) and 22A-328 (21, 24, and 28 March 2022; mean epoch 2022.227).
Both used the most extended A configuration and set Sgr~A* as the phase center because the SiO masers of interest fall within the primary beam.
Both used 3C286 for flux calibration, but 19A-310 used J1733$-$1304 for bandpass and delay calibration while 22A-328 used J1924$-$2914.  Rather than switch
between the science target field and a complex gain calibrator, the $1.0\pm0.1$~Jy Sgr~A* compact continuum was used for in-beam gain
calibration in both programs.  
While the 19A-310 program has been analyzed in \citet{paine2022}, we reprocess and reanalyze it here in a manner that is 
consistent with the legacy measurements, particularly 2014.249 (see Table \ref{tab:data}), and the treatment of 22A-328 observations described below.  

VLA 19A-310 observations spanned 2.25 hours (1.16 hours on-source) using a recording time of 2~s and two circular polarizations.  
The $v=0$, $v=1$, and $v= 2$ $J=1-0$ transitions of SiO and the $v=1$ $J=1-0$ transition of $^{29}$SiO were observed with 62.5~kHz channels,
but only the $v=1$ and $v=2$ transitions of SiO at 43.1221~GHz and 42.8206~GHz were detected.  Bandwidths were 128~MHz each, except for the $v=2$ spectral window,
which spanned 64~MHz.  
SiO-16 was not detected in either transition in the 2020 epoch, and IRS~7, IRS~9, and IRS~28 were only detected at $>5\sigma$ in the $v=1$ transition (IRS~9 $v=2$
was not observed).

VLA 22A-328 observations spanned 5 hours (3.77 hours on-source) in each of three observing sessions.  
The $v=1$ and $v=2$ $J=1-0$ transitions of SiO were observed using 3~s integrations, two circular polarizations, and 100~kHz channels
spanning 128~MHz.  
Only IRS~7 lacked a $>5\sigma$ detection in one of the transitions ($v=2$).

We used CASA \citep{CASA} for calibration, imaging, and coordinate measurements.
Prior to calibration, the data were averaged in frequency from 62.5~kHz and 100~kHz channels to 187.5~kHz (1.3~km~s$^{-1}$)
and 200~kHz (1.4~km~s$^{-1}$)
channels and in time from 2~s and 3~s records to 6~s records for the 2020 and 2022 observations, respectively.  
Using Sgr~A* for the complex gain calibration provides in-beam calibration of the masers and forces the Sgr~A* continuum to be the
phase center.  The absolute astrometry is therefore lost, but the reference frame and dynamical quantities of interest can be obtained from the relative
coordinate offsets of the masers compared to Sgr~A*, so relative astrometry is adequate for our science goals.
Sgr~A* shows an apparent 6.4 mas~yr$^{-1}$ proper motion when compared to background quasars due to the Solar orbit about the Galactic Center \citep{reid2020,xu2022}, and its position was updated for each observation.

Imaging used the CASA \texttt{tclean} algorithm centered on Sgr~A* with postage-stamp image cubes of all maser locations as ``outlier'' fields.
The outlier fields are cleaned simultaneously with Sgr~A*.  We did not subtract the continuum from the spectral line data.  Cleaning was
performed down to five times the per-channel rms noise in the dirty cubes.  All three sessions of 22A-328 were incorporated into a single spectral cube for
each SiO transition for each maser.  Figure \ref{fig:spectra} shows all spectra for the $v=1$ transition.  
The rms noise per channel was $\sim$2--3~mJy~beam$^{-1}$ in 2020 and $\sim$1~mJy~beam$^{-1}$ in 2022.

To be consistent with previous work by \citet{reid2003,reid2007}, and contrary to the $uv$-based fitting used by \citet{paine2022}, we measure
maser positions in the image plane.
We used the CASA routine \texttt{imfit} to fit 2D Gaussians to measure the centroid of the Sgr~A* continuum and each maser in every channel in each transition
independently.  Maser coordinates were obtained from a variance-weighted average of the channel-by-channel centroids with peak fluxes $>5\sigma$, incorporating both transitions.  Sgr~A* coordinates were
calculated from the variance-weighted channel centroids over the entire continuum.
Typical maser coordinate uncertainties are 0.2 milliarcsec (1.6 AU at 8.2 kpc), which is a substantial improvement over most legacy measurements by a factor of roughly 2--4.

We combined the newly-measured maser coordinates with those from the legacy observations
listed in Table \ref{tab:data} to form time series spanning up to 25.8 years.  After the astrometric corrections described below are applied
to the time series, linear fits provide proper motions.

\begin{deluxetable*}{lrrrccrcc}
  \tablecaption{\label{tab:results} SiO Maser Angular Offsets and Proper Motions}
  \tablehead{ \colhead{Name} & \colhead{$v_{\rm LSR}$} & \colhead{RA Offset\tablenotemark{a}} & \colhead{Dec.\ Offset} & \colhead{PM RA} &  \colhead{PM Dec.} & \colhead{$\chi^2_\nu$} & \colhead{Ref.\ Epoch} & \colhead{$N_{\rm Obs}$}\\
 &  (km s$^{-1}$) & \colhead{ (arcsec)} & \colhead{(arcsec)} & (mas yr$^{-1}$) &  (mas yr$^{-1}$)  } 
  \startdata
IRS 9  &   $-$341 &   $+5.71043 \pm 0.00009$  &  $-6.30688 \pm 0.00023$ &  $+3.080 \pm 0.016$ &  $+2.291 \pm 0.033$ & 1.0 & 2017.946 & 8 \\
IRS 7  &   $-$114 &   $+0.03330 \pm 0.00500$  &  $+5.49028 \pm 0.00500$  & $-0.002 \pm 0.044$  & $-4.665 \pm 0.093$ & 1.2 & 2013.582  & 6 \\
SiO-14 & $-$111  &  $-7.62578 \pm 0.00032$ &  $-28.46850 \pm 0.00046$  & $+2.073 \pm 0.041$  & $-0.969 \pm 0.064$ & 4.3 & 2017.153 & 8 \\
IRS 12N &   $-$65 &   $-3.27773 \pm 0.00013$ &   $-6.94708 \pm 0.00015$ &  $-1.122 \pm 0.021$ &  $-2.834 \pm 0.024$ & 2.8 & 2019.686 &  9 \\ 
IRS 28   &    $-$54 &  $+10.49199 \pm 0.00030$ &   $-5.86884 \pm 0.00050$ &  $+1.548 \pm 0.050$  & $-5.493 \pm 0.088$ & 2.9 &  2014.235 & 8  \\
SiO-15   &   $-$35 &  $-12.46900 \pm 0.00029$ &  $-11.06769 \pm 0.00038$ &  $-2.562 \pm 0.058$ &  $+0.738 \pm 0.068$ & 1.1 & 2017.505 & 6  \\
IRS 10EE  &   $-$28 &   $+7.68504 \pm 0.00011$  &  $+4.17765 \pm 0.00017$  & $+0.070 \pm 0.017$  & $-1.984 \pm 0.020$ & 2.3 & 2017.308 & 9  \\
IRS 15NE  &   $-$11 &   $+1.20422 \pm 0.00019$  & $+11.25164 \pm 0.00028$  & $-1.925 \pm 0.019$ &  $-5.802 \pm 0.028$ & 1.6 &  2010.230 & 9  \\
SiO-16   &    +7  & $-26.42046 \pm 0.00067$  & $-34.47238 \pm 0.00124$ &  $-0.002 \pm 0.093$ &  $-1.989 \pm 0.170$ & 16.9 & 2017.089 & 7  \\
SiO-6    &   +52  & $+35.25587 \pm 0.00106$ &  $+30.68278 \pm 0.00227$  & $+2.719 \pm 0.113$  & $+2.507 \pm 0.248$ & 11.4 & 2009.959 & 8 \\
SiO-17   &   +53  &  $+8.08338 \pm 0.00035$ &  $-27.66156 \pm 0.00065$ &  $+2.468 \pm 0.052$ &  $+2.492 \pm 0.108$ & 7.8 & 2014.935 & 7  \\
SiO-11   &   +71  &  $+1.76111 \pm 0.00078$ &  $+40.30709 \pm 0.00151$  & $+1.704 \pm 0.131$  & $+1.904 \pm 0.230$ & 46.5 & 2014.160  & 8  \\
IRS 17    &   +74  & $+13.14134 \pm 0.00090$ &   $+5.55666 \pm 0.00148$ &  $-1.073 \pm 0.165$ &  $-1.059 \pm 0.240$ & 1.8 & 2009.404 & 6 \\ 
SiO-12   &   +82  & $-18.80861 \pm 0.00086$ &  $+42.48144 \pm 0.00177$ &  $+1.086 \pm 0.177$ &  $+1.458 \pm 0.310$ & 6.9 & 2015.756  & 7  \\
IRS 19NW  &   +84  & $+14.57819 \pm 0.00033$  & $-18.47510 \pm 0.00068$  & $+1.414 \pm 0.074$ &  $-0.702 \pm 0.124$ & 15.0 & 2019.938 & 8  \\
\enddata
\tablecomments{Coordinate offsets are with respect to the Sgr~A* radio centroid at the reference epoch, which is the position variance-weighted date of the time series (see Section \ref{sec:methods}).  The LSR velocity is
  approximate; the masers show variability in their spectral peaks and velocity centroids \citep{reid2003,reid2007,paine2022}.  The $\chi^2_\nu$ statistic characterizes the joint weighted least-squares proper motion fit in both coordinates.  The coordinate offset uncertainties in IRS~7 have been
  manually adjusted to $\pm5$~mas (see Section \ref{sec:methods}).}
\tablenotetext{a}{This offset is corrected for Declination:  it is $\Delta {\rm RA} \cos({\rm Dec.})$.}
\end{deluxetable*}

\section{\label{sec:methods} Astrometric Methods}

Masers (and stars) in the vicinity of Sgr~A* may not appear to be exactly where they physically lie.  Light propagation and observer-induced effects
such as solar gravitational deflection and aberration can cause the entire field of view to shift, which is not a problem for relative coordinate measurements,
but these effects are also differential, causing relative astrometric offsets between objects as observed.  In general, any phenomenon that deflects or appears to deflect
light rays and depends on direction will be differential and therefore stretch, shear, or rotate the observed field of view.

It is important to differentiate between relative astrometric offsets from Sgr~A* that depend on the observation epoch and those that are
stable over time.  Epoch-dependent relative offsets must be determined and removed from astrometric time series to obtain proper motions.
Time-stable offsets must be quantified in order to determine the actual physical locations of stars for kinematic or dynamical modeling, such as
characterizing the metric around the Sgr~A* black hole or the mass distribution of the inner parsec \citep[e.g.,][]{gravity2022}.

Time-dependent differential astrometric offsets include aberration, terrestrial precession-nutation, and
solar gravitational deflection.  
Aberration caused by an observer's motion will be differential because its amplitude depends on direction \citep[e.g. the CMB and galaxies
show a dipole;][]{smoot1977,ellis1984,darling2022}.  The dominant contribution is the Solar motion within the Galaxy, which
produces a steady apparent motion of Sgr~A* \citep{reid2020}, but the Earth's orbit adds an aberration epicycle that does depend on the observation
epoch.  Terrestrial precession-nutation involves the secular precession of the celestial pole plus epicycles about this pole,
which are necessarily time-dependent.  Finally, the solar mass causes measurable gravitational deflection, even at large angular offsets, and the solar-Sgr~A*
angular separation depends on the observation epoch.

Corrections for time-dependent differential offsets were applied to all data, new and legacy, using \texttt{astropy.coordinates}
tools \citep{astropy:2013,astropy:2018}.  
Starting with the observed maser offsets and the Sgr~A* J2000 coordinates, we calculate the ``mean'' maser J2000 coordinates.   These coordinates
include a precession correction from the epoch of observation to J2000 but do not include the above effects from
aberration, nutation, or gravitational light deflection.
Next, we precess the maser and Sgr~A* coordinates from J2000 to the equinox of each observed epoch (``apparent'' coordinates)  and then transform to 
a precessed geocentric J2000 coordinate system.  The geocentric transformation includes the effects of aberration, the precession and nutation of the Earth's rotation axis, and gravitational deflection of incoming rays \citep{iau2005}.  Finally, we subtract the precessed and
transformed Sgr~A* coordinates from the precessed and transformed maser coordinates to obtain a relative maser offset.
To correct for time-dependent differential astrometric offsets we find the difference between the coordinate offsets obtained from
the above transformations and the J2000 coordinate offsets as observed.  This difference is then subtracted from the observed coordinate offsets.  Figure \ref{fig:offsets} in Appendix \ref{sec:offsets} shows an example of the corrections for one epoch.
Time-independent offsets, such as the solar-Galactic Center aberration, are not removed by this process.
A similar process, using different software, was applied to the SiO maser astrometry used in \citet{sakai2019} but not to the \citet{reid2007} results.  

These differential corrections generally slightly reduce the scatter in the residual time series after fitting for proper motions (described below).
This is encouraging, and suggests that the process is providing reasonable time-dependent astrometry.
However, the corrections are typically smaller than the variation in the astrometry, and the
  proper motions and reference frame stability are not significantly altered compared to the no-corrections case.
The magnitude of the differential corrections scales linearly with angular separation from Sgr A* in a given epoch and varies from epoch to epoch.
The corrections range from $\sim$0.1~mas to 4.3~mas in absolute value and are similar to the astrometric
uncertainty in each coordinate in each epoch, except for masers with large offsets from Sgr A*.  The latter have corrections larger
than centroid uncertainties due to the linear scaling of the corrections.  
Table \ref{tab:time_series} in Appendix \ref{sec:offsets} lists the full astrometric time series and the differential corrections for all masers in all epochs.

IRS~7 requires special treatment:  it has supergiant luminosity, its SiO maser emission distribution may span 10~mas, and its $v=1$ $J=1-0$ maser shows
substantial variability,
both in flux density and in velocity \citep{reid2003,reid2007}.  To wit, the $J=1-0$ maser decreased in brightness by a factor of 8, and the 
$-$124~km~s$^{-1}$ maser component faded below the flux density of the $-$103~km~s$^{-1}$ component
from 1995.49 to 2000.85.  The dominant component at $-$103~km~s$^{-1}$ persists through the current epoch (Figure \ref{fig:spectra}),  but the
$J=2-1$ transition resembles the pre-2000 $J=1-0$ spectrum:
it is about 10 times brighter and peaks at roughly $-$123~km~s$^{-1}$ \citep{paine2022}.
The ALMA $J=2-1$ astrometry in 2015.27 and 2017.72, however,  is statistically
consistent with temporally bracketing VLA $J=1-0$ astrometry in 2014.18 and 2020.99, which
is at odds with the possibility of a shift of the $v=1$ $J=1-0$ maser emission from one side of the supergiant to the other.  We conclude that for the purposes of a current and near-future reference frame determination,
the proper motion and position of IRS~7 should rely on the last 20 years of observations and omit those made before the dramatic change in the $J=1-0$ emission.  The coordinates and proper motions
presented in Table \ref{tab:results}
and Figure \ref{fig:pms} rely on epochs 2006.200--2022.227, and the coordinate uncertainties in the
astrometric solution have been set to $\pm$5~mas to allow for the likely maser offsets from the stellar photocenter,
following \citet{reid2007} and \citet{paine2022}.
Table \ref{tab:time_series} includes the omitted epochs for posterity.

Proper motion measurements in each coordinate require a linear fit to the now-corrected offset coordinate time series.
The offset position $(\Delta \alpha \cos \delta,\Delta \delta)$ of a given stellar maser with respect to Sgr~A* observed in epoch $t_{\rm obs}$  is
\begin{eqnarray}
  \Delta \alpha \cos \delta & = &\Delta \alpha_{\rm ref} \cos \delta_{\rm ref}+ \mu_\alpha (t_{\rm obs} - t_{\rm ref})\label{eqn:dra}, \\
  \Delta \delta & = &\Delta \delta_{\rm ref} + \mu_\delta (t_{\rm obs} - t_{\rm ref}) \label{eqn:ddec}
\end{eqnarray}
for offsets $\Delta \alpha_{\rm ref}$ and $\Delta \delta_{\rm ref}$ at reference epoch $t_{\rm ref}$ given proper motions $\mu_{\alpha,\delta}$.
We do not include curvature in the fits, which would
correspond to acceleration (see \citealt{paine2022} for that analysis and limits on accelerations).  After exploring several fitting methods, described
in Appendix \ref{sec:pms}, we chose the simplest:  linear variance-weighted least-squares fits using a reference epoch as the intercept of the linear fit.
The reference epoch $t_{\rm ref}$ is the variance-weighted mean date in the time series, where the error variance is the sum in quadrature
of the coordinate uncertainties of each epoch.  The ``intercept'' of each proper motion fit to the coordinate time series refers to the
coordinates of the maser at the reference epoch.  This method generally shows negligible correlation between the slope and intercept of the
linear fit.  The proper motion parameters also show negligible correlation between RA and Dec., which are fit
jointly.  See Appendix \ref{sec:pms} for details.

\begin{figure}[t]
\begin{centering}
  \includegraphics[scale=0.48,trim= 20 20 20 20,clip=false]{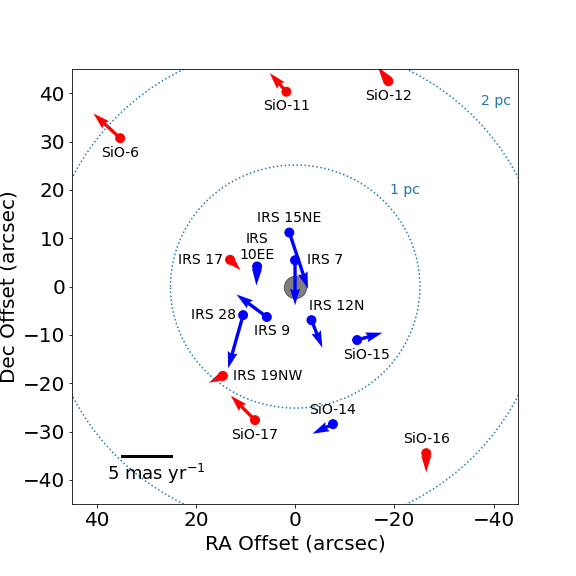}
  \caption{Locations in epoch 2022.227 and proper motions of the stellar SiO masers used for the Galactic Center astrometric reference frame (Table \ref{tab:results} and Figure \ref{fig:pms}).
    The color indicates the sign of the stellar radial velocities (red is positive; blue is negative). The grey circle marks Sgr~A* (not to scale).
   Projected distances assume a Galactic Center distance of 8.2 kpc, and the RA offset is corrected for Declination.}
\label{fig:map}
\end{centering}
\end{figure}

\section{\label{sec:results} Results}

Table \ref{tab:results} lists the LSR velocity of each maser (see \citealt{paine2022} for a detailed study),
coordinate offsets from Sgr A* at the reference epoch for each maser,  the proper motions in each coordinate, a reduced chi-squared
statistic for the joint proper motion fit,
the reference epoch, and the number of epochs used for the fits.  Figure \ref{fig:map} shows an overview of the maser locations and proper motions and
indicates whether masers are redshifted or blueshifted along the line of sight.  Figure \ref{fig:pms} shows the coordinate time series,
proper motions, and linear fit residuals for each maser.   To obtain the coordinate offsets from Sgr~A*
for a given stellar maser at a specific time, one would employ
the parameters presented in Table \ref{tab:results} in Equations \ref{eqn:dra} and \ref{eqn:ddec}.  

\begin{figure*}[t]
\begin{centering}
  \includegraphics[scale=0.30,trim= 20 20 20 0,clip=false]{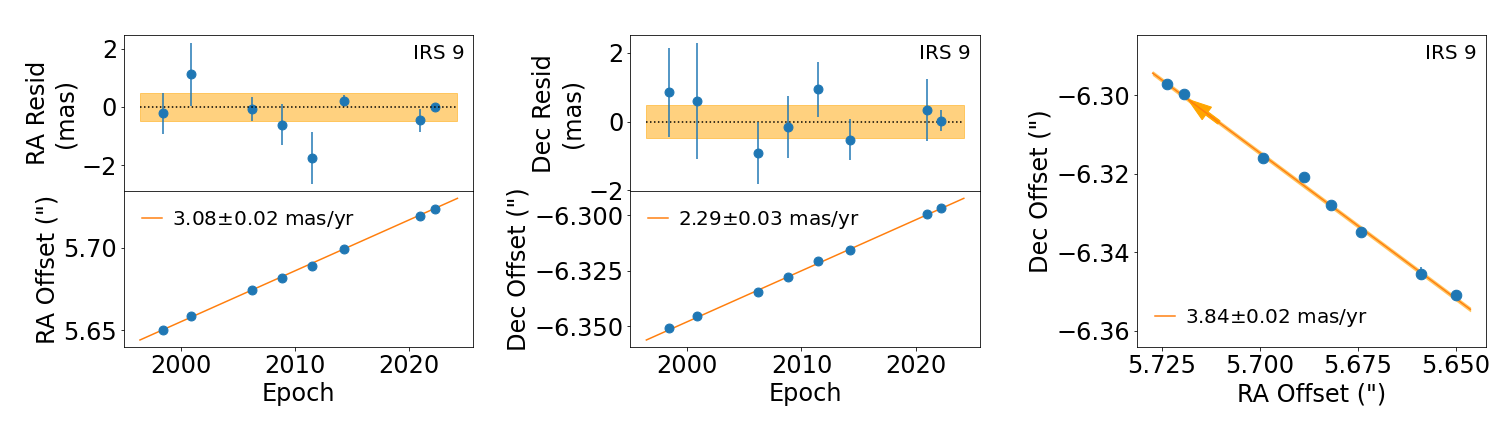}
  \includegraphics[scale=0.30,trim= 20 20 20 0,clip=false]{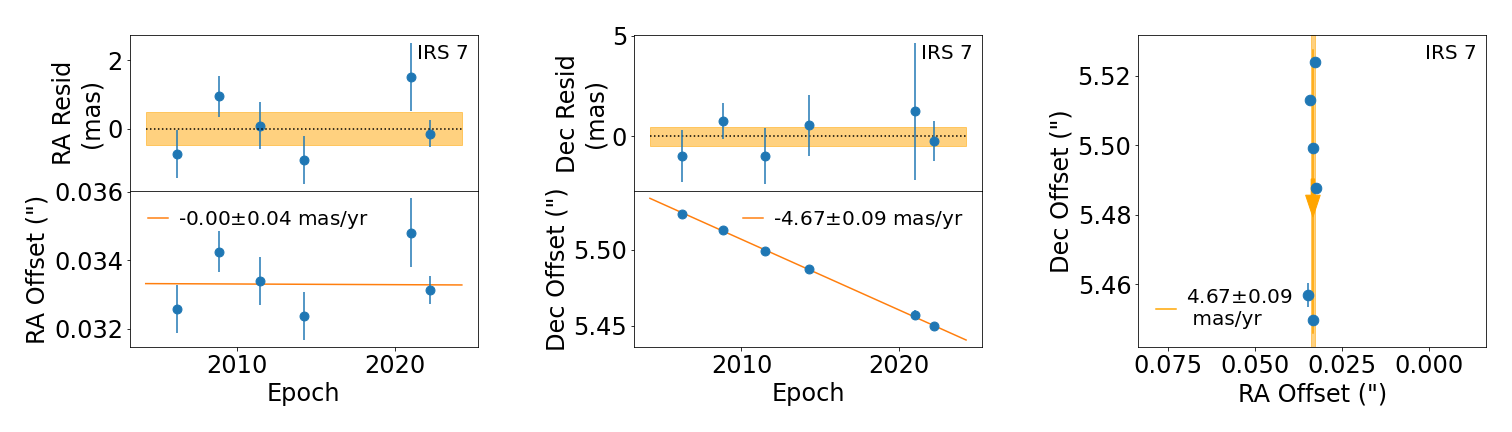}
  \includegraphics[scale=0.30,trim= 20 20 20 0,clip=false]{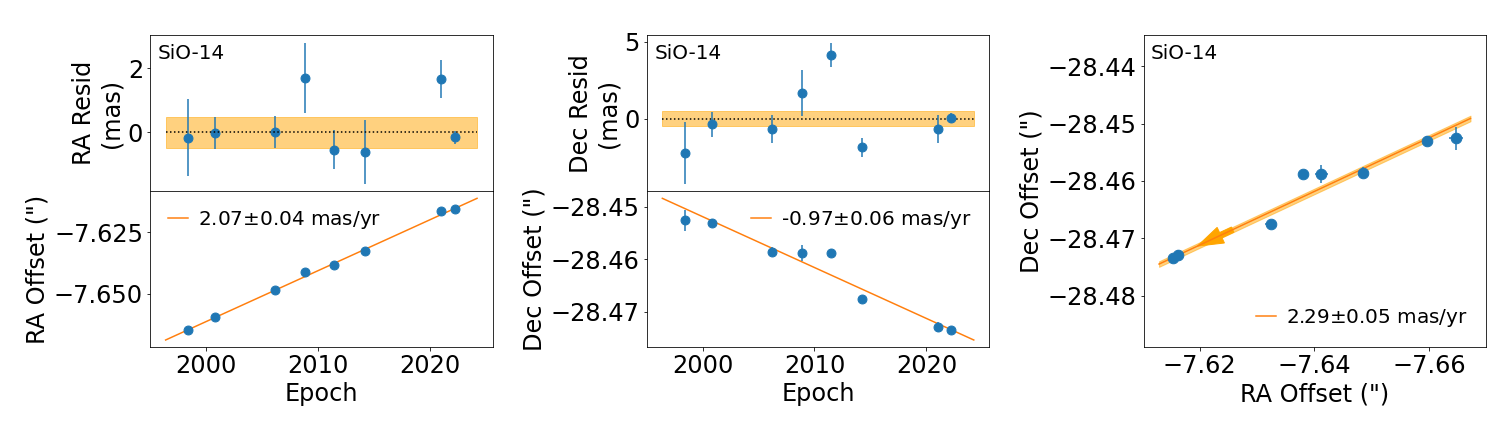}
  \includegraphics[scale=0.30,trim= 20 20 20 0,clip=false]{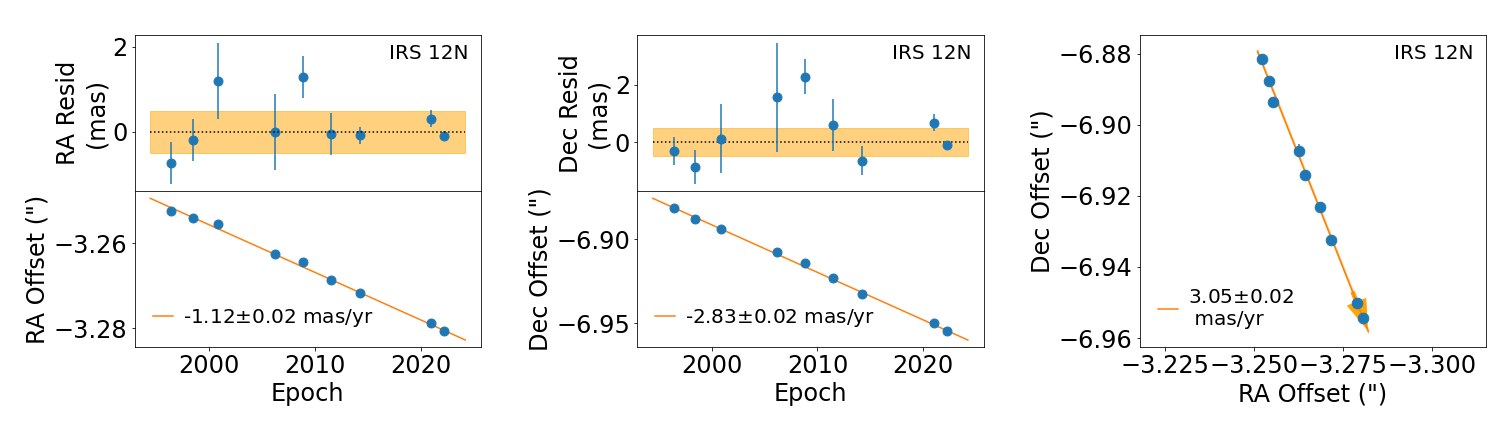}
  \includegraphics[scale=0.30,trim= 20 20 20 0,clip=false]{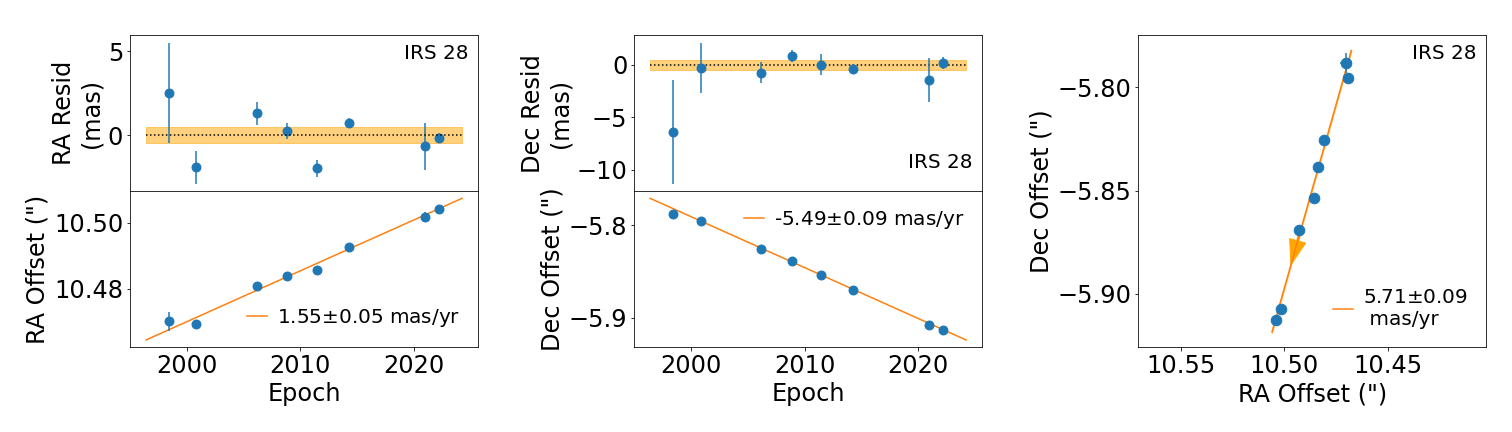}
  \caption{Left and center columns: Time series in each coordinate for each stellar maser.  The lower panel for each shows
    the coordinate offset from Sgr~A* and the weighted least-squares linear proper motion fit.  The upper panel shows the best-fit
    residual versus epoch. The shaded region indicates $\pm4$~AU.
      RA offsets are true angular offsets (i.e., they are corrected for $\cos({\rm Dec.})$). Right column:  Sky
      tracks for the masers.  The arrows indicate the direction of the 2D proper motion. }
\label{fig:pms}
\end{centering}
\end{figure*}

\addtocounter{figure}{-1}

\begin{figure*}[h!t]
\begin{centering}
  \includegraphics[scale=0.30,trim= 20 20 20 0,clip=false]{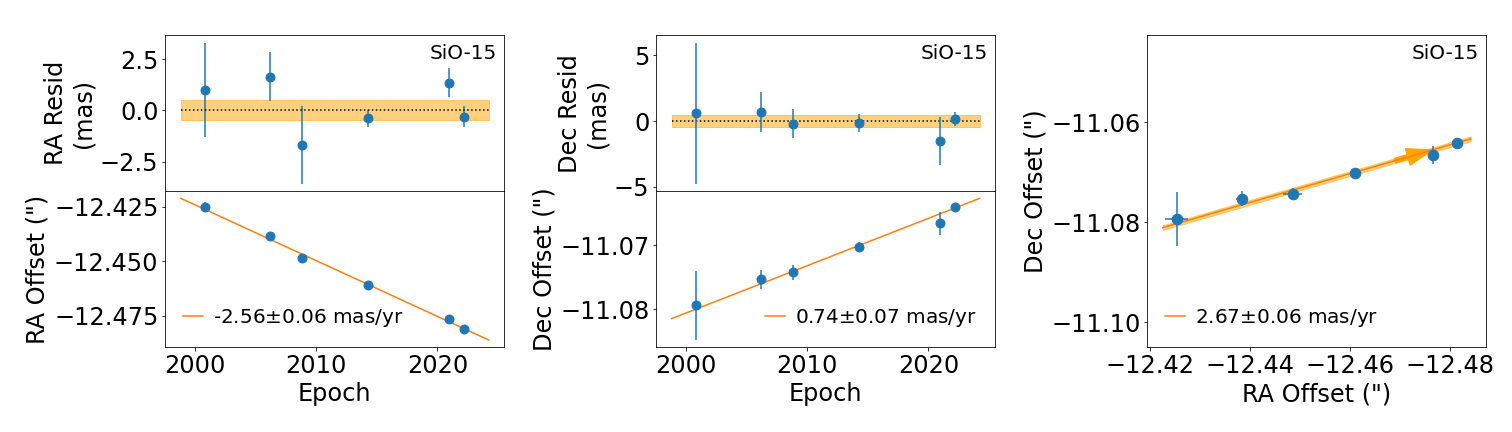}
  \includegraphics[scale=0.30,trim= 20 20 20 0,clip=false]{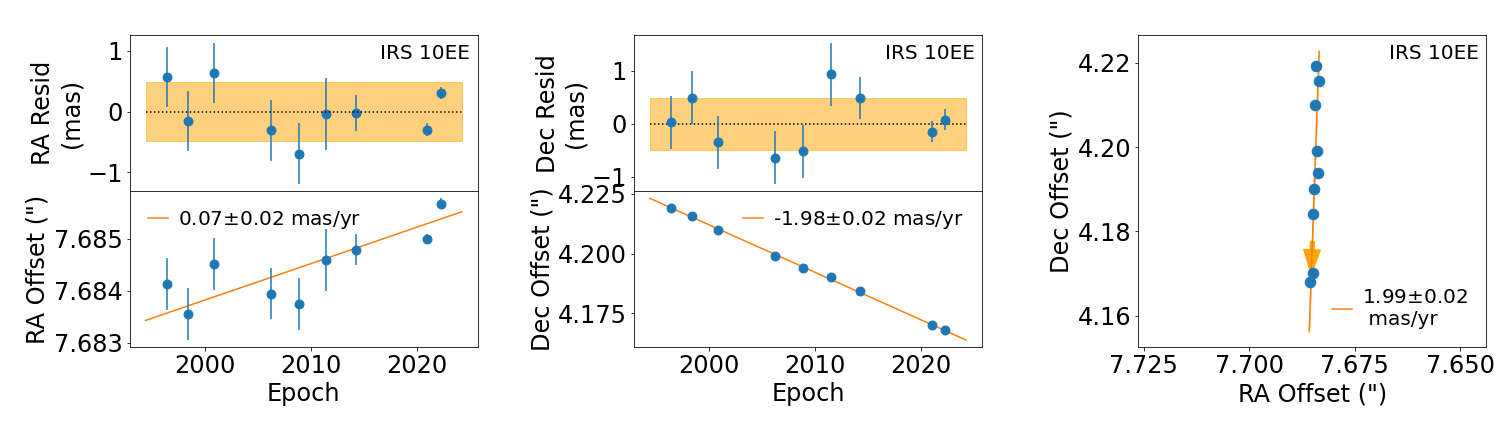}
  \includegraphics[scale=0.30,trim= 20 20 20 0,clip=false]{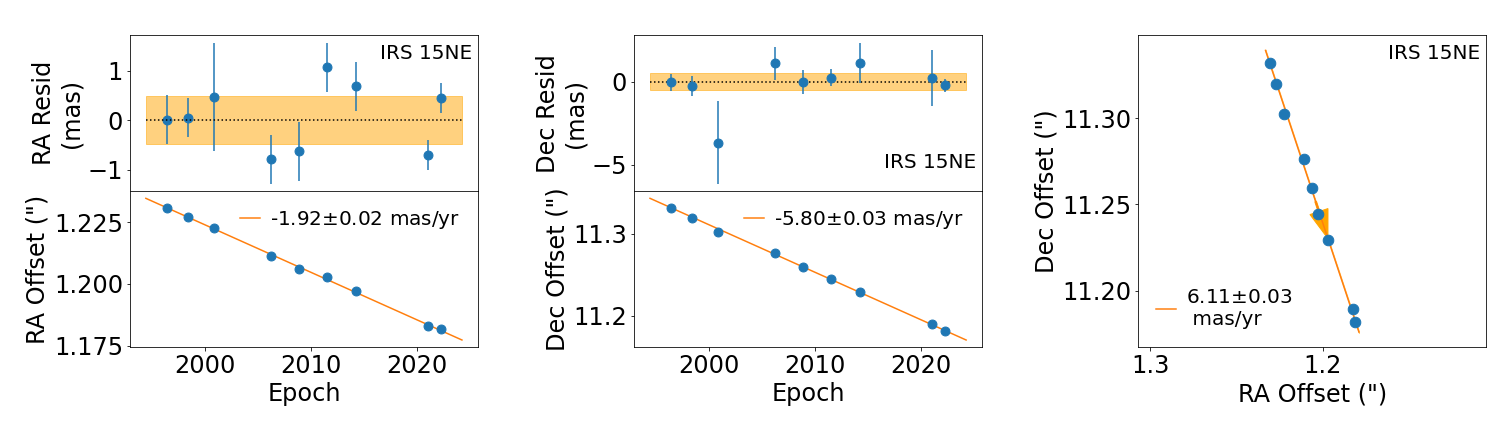}
  \includegraphics[scale=0.30,trim= 20 20 20 0,clip=false]{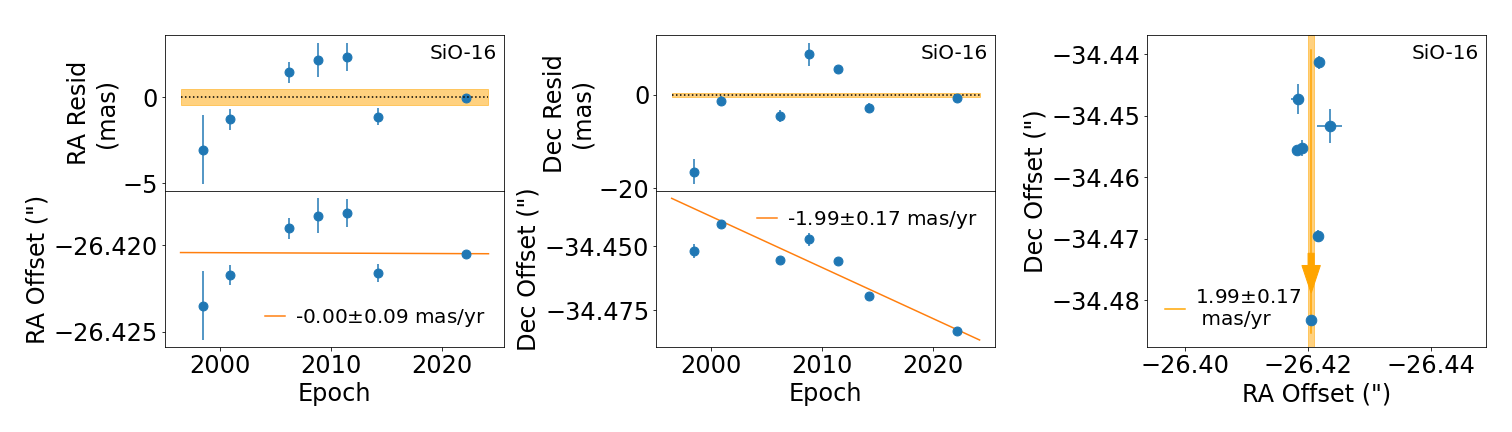}
  \includegraphics[scale=0.30,trim= 20 20 20 0,clip=false]{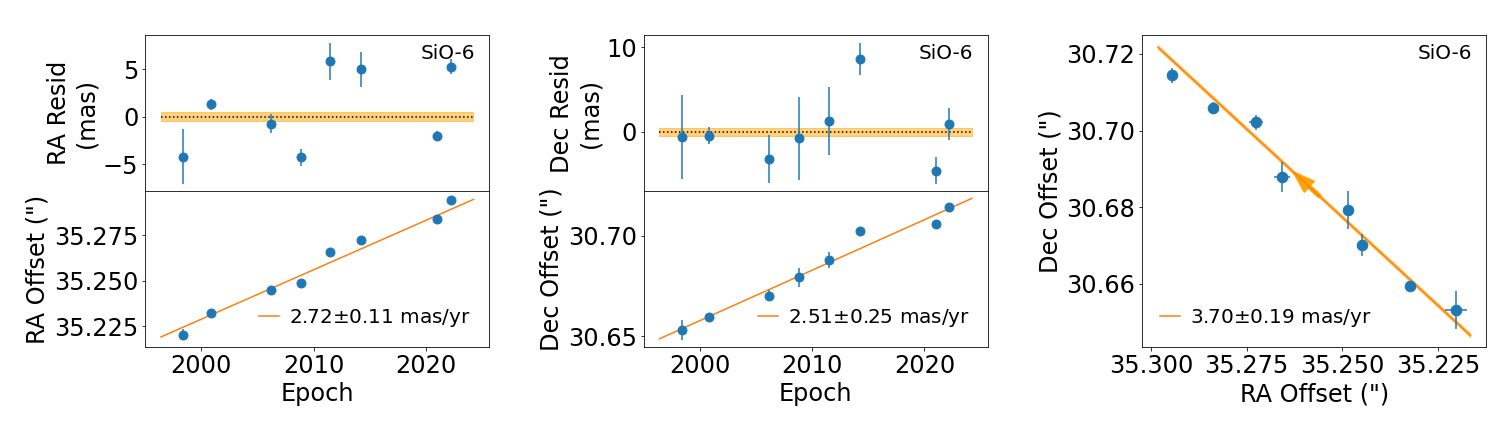}
  \caption{{\it continued}}
\end{centering}
\end{figure*}


\addtocounter{figure}{-1}

\begin{figure*}[h!t]
\begin{centering}
  \includegraphics[scale=0.30,trim= 20 20 20 0,clip=false]{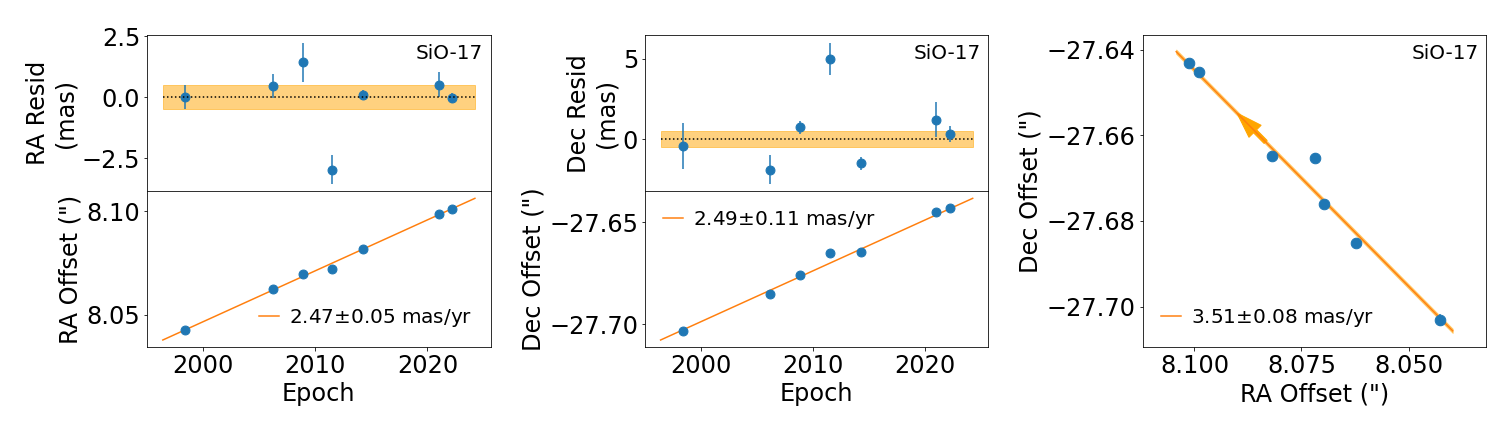}
  \includegraphics[scale=0.30,trim= 20 20 20 0,clip=false]{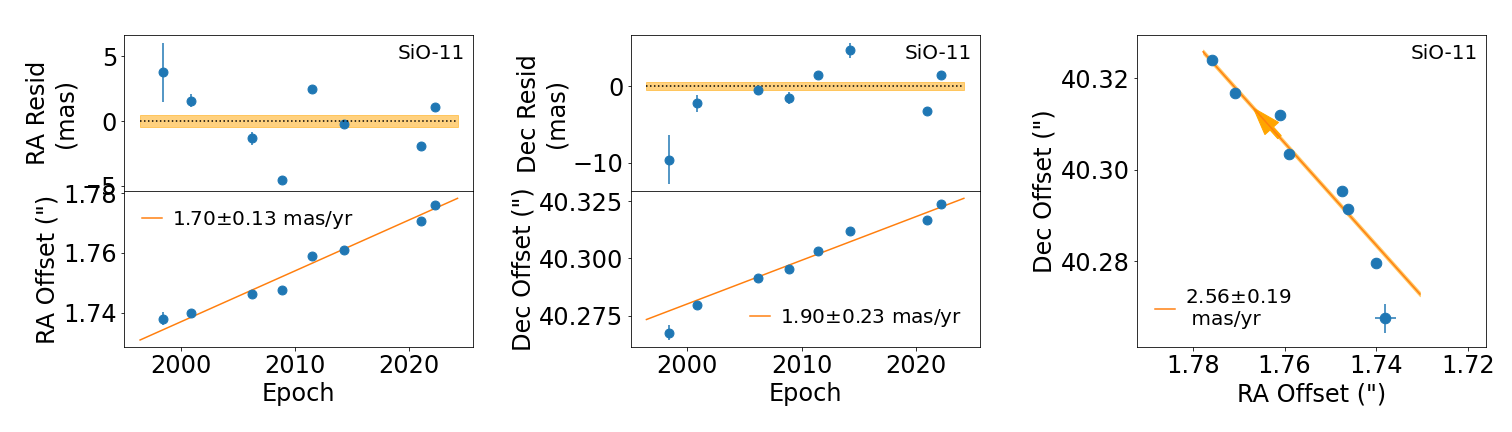}
  \includegraphics[scale=0.30,trim= 20 20 20 0,clip=false]{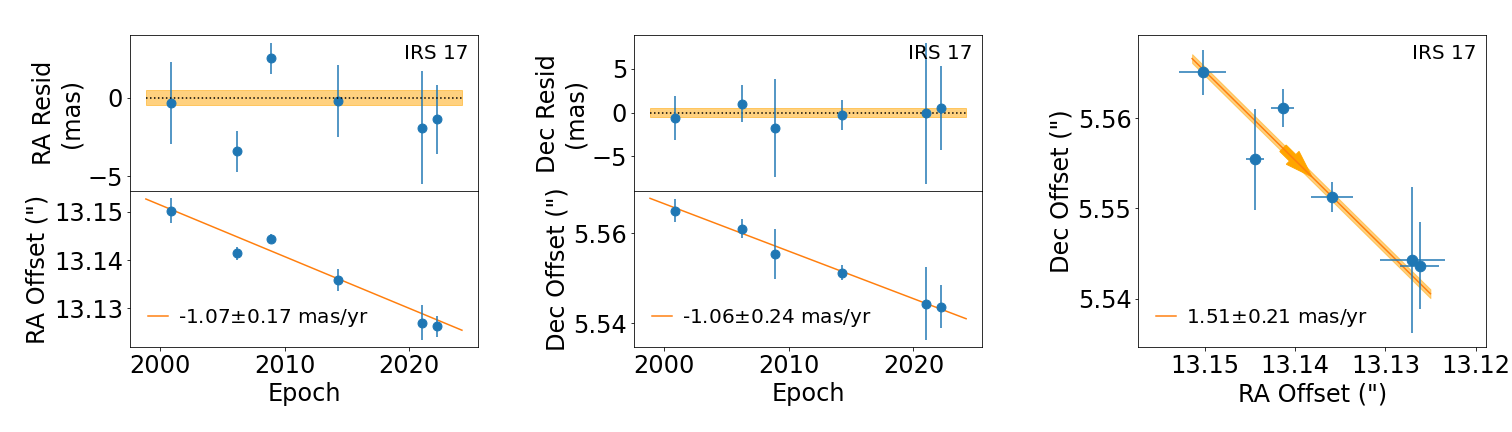}
  \includegraphics[scale=0.30,trim= 20 20 20 0,clip=false]{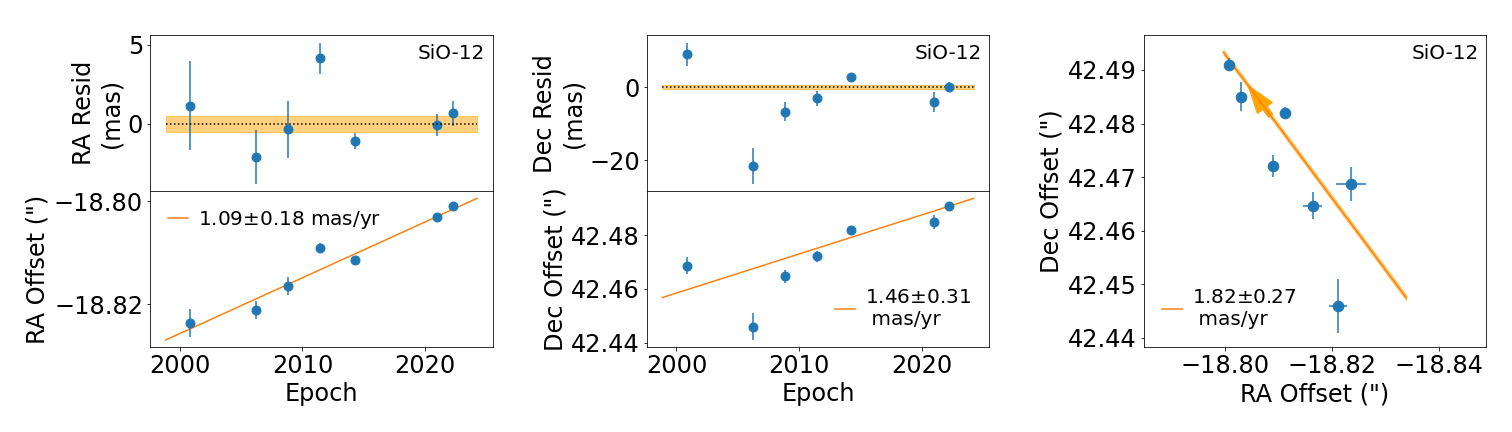}
  \includegraphics[scale=0.30,trim= 20 20 20 0,clip=false,]{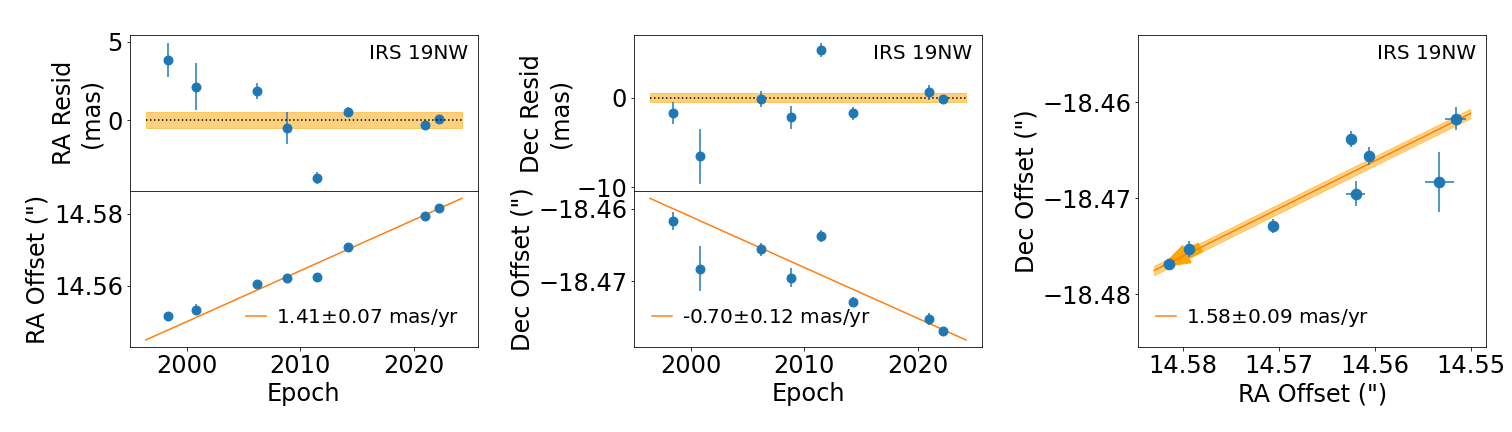}
  \caption{{\it continued}}
\end{centering}
\end{figure*}

\clearpage

Most masers have 7--9 epochs in their time series (SiO-15 and IRS~17 have 6, and for IRS~7 we use 6),
and the post-2000 epochs tend to include more masers.  The
per-epoch uncertainty in coordinates ranges from 0.1~mas to 8.1~mas,
with uncertainties uniformly smaller for RA than for Dec.\ due to a north-south
elongated synthesized beam.  Uncertainties tend to be larger in earlier epochs (less sensitivity and shorter integration times)
and for fainter masers (lower signal-to-noise).
Table \ref{tab:time_series} lists the coordinates for each maser in every epoch.  
The mean uncertainty in the offset coordinate of the masers in the reference epoch is 0.46~mas in RA and 0.84~mas in Dec., excluding IRS~7 (see Section \ref{sec:methods}).
These correspond to 3.7 AU and 6.9 AU, respectively, at 8.2~kpc.

The formal uncertainties in the measured position of Sgr~A* are 1~$\mu$as and 2~$\mu$as in
RA and Dec.\ in epoch 2022.227 and are therefore negligible compared to the uncertainties in the maser coordinates.
It is important to note that these very small uncertainties in the Sgr~A* position are strictly statistical, and the coordinates of Sgr~A* are assigned {\it a priori} because
it is the phase center used for complex gain calibration.

However, recent work by \citet{xu2022} found a $\sim$30~mas offset from the canonical absolute position of Sgr A* \citep{reid2020}.
This causes second-order astrometric offsets in the maser positions of order $30\ {\rm mas} \times\Delta \theta$ where $\Delta \theta$,
expressed in radians, is the angular offset from Sgr A*.  For the masers presented here, the error is roughly 1--7 $\mu$as.  This
is negligible compared to other uncertainties and systematics, and therefore no correction was applied to the astrometry.

The mean maser proper motion uncertainties are 0.07~mas~yr$^{-1}$ in RA and 0.12~mas~yr$^{-1}$ in Dec., corresponding to 2.7~km~s$^{-1}$ and 4.6~km~s$^{-1}$.
As seen in Figure \ref{fig:pms}, the residuals from the linear time series fits often have
significant outliers.  These outliers are not consistent across all masers at a fixed epoch
(i.e., there do not seem to be epochs with bad astrometry), and they are not restricted to particular stars
or coordinate directions.  Residuals are often within roughly 1--2~mas (Figure \ref{fig:pms}); 1~mas
corresponds to $\sim$8~AU, which is the typical size of SiO maser distributions around evolved stars \citep{cotton2008}.
Residuals can, however, be as large as $\sim$5~mas or $\sim$40~AU (Figure \ref{fig:pms}).
The nature of the variation in residuals remains unknown, but suggests a systematic effect that should be addressed in future work.  It is
clear, however, from these long time baselines that astrometric trends (i.e., proper motions) can be measured
despite substantial single-epoch departures.  

Among the best-measured masers are the bright ones:  IRS~9, IRS~12N, and IRS~10EE, which reach coordinate uncertainties of
0.09--0.13~mas (0.7--1.1~AU) in RA and 0.15--0.23~mas (1.2--1.9~AU) in Dec.  These uncertainties are smaller than the expected size of the
maser-emitting regions in the stellar atmospheres.  The proper motions of these masers have uncertainties of 0.016--0.021~mas~yr$^{-1}$ (0.6--0.8~km~s$^{-1}$) in RA and
0.021--0.033~mas~yr$^{-1}$ (0.8--1.3~km~s$^{-1}$) in Dec., showing that it is possible to reach sub-km~s$^{-1}$ precision in measurements of
transverse velocity \citep[also demonstrated by][]{paine2022}.

The reference frame stability, the uncertainty in the variance-weighted mean proper motion of the maser ensemble,
is 8~$\mu$as~yr$^{-1}$ in RA and 11~$\mu$as~yr$^{-1}$ in Dec.\ or 0.30~km~s$^{-1}$ in RA, 0.44~km~s$^{-1}$
and in Dec. This sub-km~s$^{-1}$ measurement is 2.3 times smaller than the previous value \citep{sakai2019}
and represents a new benchmark for the maser-based reference frame.  This new reference frame stability is in agreement
  with the predictions made by \citet{yelda2010} and \citet{sakai2019} and should enable observation of the
  apocenter shift of the star S0-2 caused by relativistic prograde precession (Schwarzschild precession; \citealt{weinberg2005}).  It should be noted that this general relativistic effect was detected for S0-2 (aka S2)
  by the \citet{gravity2020}, but the precision-limiting factor in the measurement was the radio-to-infrared
  reference frame conversion of \citet{plewa2015}.

\begin{figure}[t]
\begin{centering}
  \includegraphics[width=0.48\textwidth,trim=20 20 0 0,clip=true]{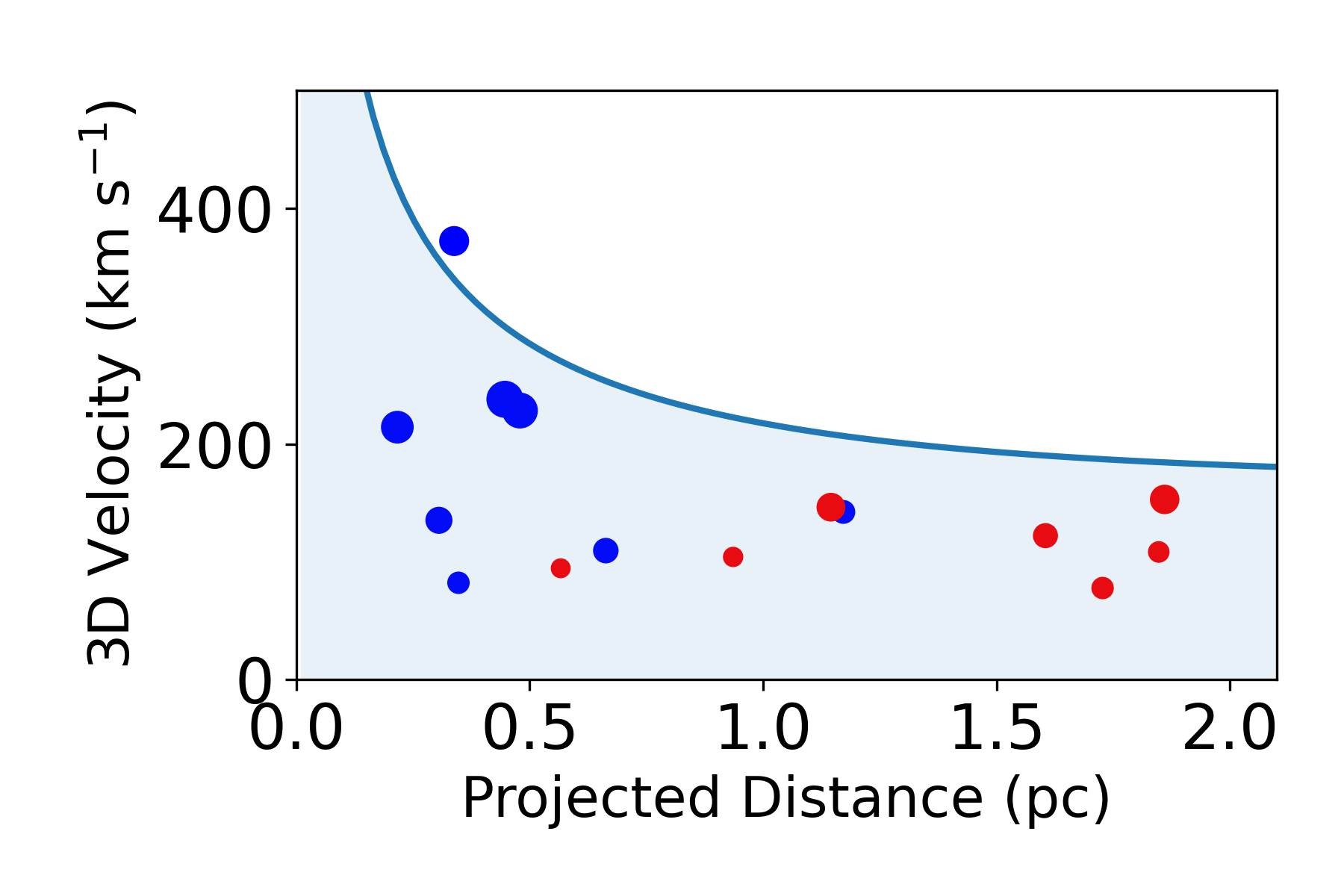}
  \caption{Three-dimensional velocity of stellar masers versus projected distance from Sgr~A*.  Point color indicates the sign
    of the radial velocities (red is redshifted, blue is blueshifted).  The size of the points scales
    linearly with the transverse velocity, spanning 59--241~km~s$^{-1}$.  Velocity errorbars are uniformly
    smaller than the data points.  The blue line indicates the
    upper bound on 3D velocity based on the enclosed stellar and black hole masses (see text and Equation \ref{eqn:v3d}).
   Projected distances assume a Galactic Center distance of 8.2 kpc.}
\label{fig:v3d}
\end{centering}
\end{figure}


\section{\label{sec:discussion}Discussion}

The astrometry in this and previous work relies on fitting Gaussian brightness distributions to planes in maser image-velocity cubes.
In contrast, \citet{paine2022} uses $uv$-based fitting, often of several masers simultaneously.  The per-epoch astrometry is generally in agreement between the two methods, including
for the 2020.988 epoch that is included in both studies.
The derived proper motions also show good agreement, although the \citet{paine2022} time baseline is
shorter.  However, this study utilized additional epochs, some of which provided 86~GHz maser-based positions.  

For many masers --- but not all --- the scatter about the linear proper motion fit is larger than the
formal uncertainties would suggest;  i.e., $\chi^2_\nu \gg 1$ (Table \ref{tab:results}).  
That this is not consistently true for all masers suggests that there is no consistent systematic
effect influencing the astrometry, and there do not appear to be specific outlier epochs.  
Possible explanations for the offsets include physical and instrumental effects, but it is difficult to identify
systematics that could produce the observed magnitude of the offsets that are not consistent across all
masers or limited to specific epochs.  Stellar winds, pulsation, maser variability, and stellar
companions are possible sources of
real offsets in SiO masing regions, but these are unlikely to produce the few-mas single-epoch departures from the observed proper motion trends.  One mas is equivalent to 8.2~AU, roughly the diameter of the stellar maser-emitting regions.  Instrumental or calibration systematics
should generally affect all masers in a given epoch, and might scale with distance from Sgr~A*.  It is noteworthy
that the masers with the highest $\chi^2_\nu$ values are all redshifted and generally at the largest separations from Sgr~A*.  

Regardless of the source of the astrometric variation, one could examine the magnitude and impact of
  an intrinsic scatter added in quadrature to the measurement uncertainties.
  In Appendix \ref{sec:errors}, we examine
  expanded uncertainties in the time series and find that while larger uncertainties are favored to fit a linear
  secular trend model in maser offsets from Sgr A* for 60\% of the maser coordinates, the resultant
  proper motions are formally consistent with those obtained from the weighted least-squares fits using the
original measurement uncertainties.  


\section{\label{sec:analysis}Analysis}

Given the mass interior to the projected distance from Sgr~A*, $M_{\rm encl}$, the 3D velocity of a bound orbit has an upper limit
\begin{equation}
  v \leq \left({2 G M_{\rm encl} \over r_{\rm proj}}\right)^{1/2}. \label{eqn:v3d}
\end{equation}
The enclosed mass is the sum of the black hole mass, combined stellar mass, and any other constituents such
as gas and dark matter.  In Figure \ref{fig:v3d}
we compare the measured 3D velocities to this upper bound assuming 
$M_{\rm BH} = 4.3\times10^6$~$M_\odot$ \citep{gravity2022} and the maximal stellar mass at 1~pc
described by \citet{schodel2018}.  All stars except IRS~9, which may be unbound \citep{reid2007},
lie below this locus, in agreement with the mass limits obtained by \citet{paine2022}.
It is interesting that the blue-shifted masers tend to be closer in projection to Sgr~A* than the redshifted masers, although the transverse velocity vectors do not show preferential radial or azimuthal trends (Figure \ref{fig:map}).
3D velocities trend larger with smaller projected radius, as one would expect.

\section{\label{sec:conclusions}Conclusions}

Using new and legacy VLA observations, we have updated the SiO stellar maser astrometric reference frame
relative to the Sgr~A* 43~GHz radio continuum.   Much of the astrometry represents new benchmarks
in precision, including sub-km~s$^{-1}$ measurements of transverse velocity for some masers and
$\sim$10~$\mu$as~yr$^{-1}$ reference frame stability.  There are, however, significant single-epoch
coordinate outliers from proper motion trends for many masers that remain unexplained but provide
opportunities to further improve the astrometry if the systematic effects can be quantified and corrected.  
We have also demonstrated the value of continued and higher cadence maser monitoring.

\acknowledgments
JD and JP acknowledge support from NSF grant AST-1908122.  SS and AG acknowledge support from the
Gordon \& Betty Moore Foundation and NSF grant AST-1909554.  We thank the anonymous referee for helpful feedback.
  This research made use of NumPy \citep{NumPy}, Matplotlib \citep{Matplotlib}, and
  Astropy\footnote{\url{http://www.astropy.org}}, a community-developed core Python package for Astronomy \citep{astropy:2013, astropy:2018}.

  \facility{VLA, VLBA}
  \software{CASA \citep{CASA}, astropy \citep{astropy:2013, astropy:2018}, NumPy \citep{NumPy}, Matplotlib \citep{Matplotlib},
    \texttt{lmfit} \citep{newville2021}, \texttt{emcee} \citep{emcee}, CARTA \citep{carta2021} }


\bibliography{ms}

\appendix

\restartappendixnumbering

\section{Differential Astrometric Offsets}\label{sec:offsets}

\begin{figure}[t]
\begin{centering}
  \includegraphics[scale=0.48,trim= 20 20 20 20,clip=false]{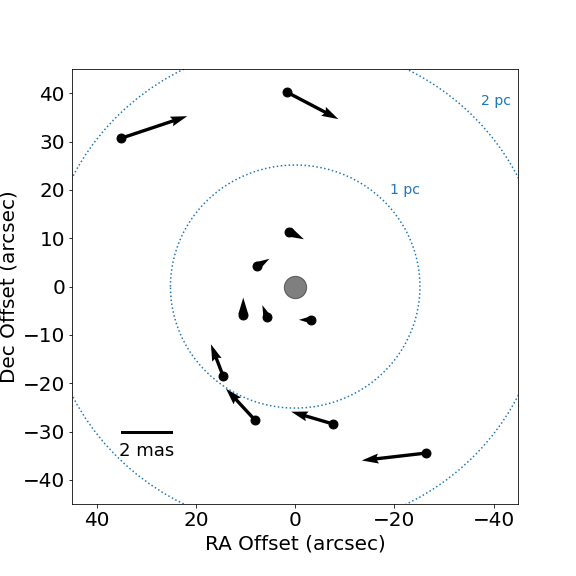}
  \caption{Example differential coordinate offsets of the stellar SiO masers used for the Galactic Center astrometric reference frame.
    This is epoch 1998.410.
    The vector field includes rotation and a radial component.  The amplitude scales linearly with angular distance from Sgr~A*.  
    The grey circle marks Sgr~A* (not to scale).  Projected distances assume a Galactic Center distance of 8.2 kpc, and the RA offset is corrected for Declination.}
\label{fig:offsets}
\end{centering}
\end{figure}

As discussed in Section \ref{sec:methods}, we correct each astrometric epoch for
differential aberration, terrestrial precession-nutation, and solar gravitational deflection.  
The net effect of these time-dependent differential astrometric corrections are offset vectors
$(\Delta$RA$\,\cos({\rm Dec.})$, $\Delta$Dec.$)$ with amplitudes
that depend linearly on the angular distance from Sgr~A* and represent a superposition of radial offsets and rotation.
For example, Figure \ref{fig:offsets} shows the offset vectors for the 1998.410 epoch.
The vector amplitudes are not constant in time but always scale as a group for each epoch.  Table \ref{tab:time_series}
lists the corrections  that have been applied to the offsets from Sgr~A* for each maser in each
epoch.

\startlongtable
\begin{deluxetable*}{lrRRrr}
  \tablecaption{\label{tab:time_series} SiO Maser Coordinate Time Series and Differential Corrections}
  \tablehead{ \colhead{Name} & \colhead{Epoch} & \colhead{RA Offset\tablenotemark{a}} & \colhead{Dec.\ Offset} & \colhead{$\Delta$RA\tablenotemark{a}} &  \colhead{$\Delta$Dec.} \\
    &  & \colhead{ (arcsec)} & \colhead{(arcsec)} & (mas) &  (mas)  } 
  \startdata
IRS 9  & 1998.410  &  5.6500 $\pm$  0.0007 &  $-$6.3508 $\pm$ 0.0013 &    0.2  &      0.5\\
          & 2000.850  &  5.6589 $\pm$  0.0011 &  $-$6.3454 $\pm$ 0.0017 &    0.1  &$-$0.7\\
          & 2006.200  &  5.6742 $\pm$  0.0004 &  $-$6.3347 $\pm$ 0.0009 &$-$0.6&      0.6\\
          & 2008.860  &  5.6818 $\pm$  0.0007 &  $-$6.3279 $\pm$ 0.0009 &    0.1  & $-$0.6\\
          & 2011.470  &  5.6887 $\pm$  0.0009 &  $-$6.3208 $\pm$ 0.0008 &     0.4 &      0.3\\
          & 2014.249  &  5.6993 $\pm$  0.0002 &  $-$6.3159 $\pm$ 0.0006 &    $-$0.5 &     0.7\\
          &  2020.988 &  5.7193 $\pm$  0.0004 &  $-$6.2996 $\pm$ 0.0009 &    $-$0.4 &    $-$0.2\\
          &  2022.227 &  5.7236 $\pm$  0.0001 &  $-$6.2971 $\pm$ 0.0003 &    $-$0.5 &     0.7\\
          \hline
IRS 7\tablenotemark{b}  &   1998.410 &  0.0378  $\pm$  0.0043 &  5.5495  $\pm$  0.0014 &    $-$0.3 &    $-$0.2\\
&  2000.850 &  0.0342 $\pm$  0.0016 &  5.5414 $\pm$  0.0030 &     0.2 &     0.4\\
&  2006.200 &  0.0326  $\pm$  0.0007 &  5.5237 $\pm$   0.0013 &     0.0 &    $-$0.5\\
&  2008.860 &  0.0343  $\pm$  0.0006 &  5.5131  $\pm$  0.0009 &     0.2 &     0.3\\
&  2011.470 &  0.0334 $\pm$  0.0007 &  5.4992 $\pm$   0.0014 &    $-$0.3 &     0.0\\
&  2014.249 &  0.0324  $\pm$  0.0007 &  5.4877  $\pm$  0.0015 &    $-$0.1 &    $-$0.5\\
&  2020.988 &  0.0348  $\pm$  0.0010 &  5.4570 $\pm$   0.0034 &     0.3 &    $-$0.1\\
&  2022.227 &  0.0331  $\pm$  0.0004 &  5.4497  $\pm$  0.0010 &    $-$0.0 &    $-$0.5\\
\hline
SiO-14 &   1998.410 & $-$7.6648 $\pm$   0.0012  &$-$28.4526 $\pm$  0.0020    &  1.7    &  0.5\\
&  2000.850 & $-$7.6596  $\pm$  0.0005 & $-$28.4531 $\pm$  0.0008    & $-$1.7    & $-$1.5\\
&   2006.200 & $-$7.6485  $\pm$  0.0005&  $-$28.4585 $\pm$  0.0009    &  0.7    &  2.8\\
&   2008.860 & $-$7.6413  $\pm$  0.0011&  $-$28.4588 $\pm$  0.0015    & $-$1.7    & $-$1.4\\
&   2011.470 & $-$7.6381 $\pm$   0.0006&  $-$28.4588 $\pm$  0.0008    &  1.5    & $-$0.6\\
&   2014.249 & $-$7.6324 $\pm$   0.0010&  $-$28.4675 $\pm$  0.0006    &  1.1    &  2.6\\
&   2020.988 & $-$7.6162 $\pm$   0.0006&  $-$28.4729 $\pm$  0.0009    & $-$1.4    &  0.9\\
&   2022.227 &  $-$7.6154 $\pm$   0.0002&  $-$28.4734 $\pm$  0.0003    &  0.9    &  2.8\\
\hline
IRS 12N &    1996.413 & $-$3.2523 $\pm$   0.0005 & $-$6.8814 $\pm$   0.0005 &     0.4 &     0.0\\
&  1998.410 & $-$3.2541  $\pm$  0.0005 & $-$6.8876  $\pm$  0.0006 &     0.5 &     0.0\\
&  2000.850 & $-$3.2554  $\pm$  0.0009 & $-$6.8936  $\pm$  0.0012 &    $-$0.5 &    $-$0.3\\
&  2006.200 & $-$3.2626  $\pm$  0.0009 & $-$6.9073  $\pm$  0.0019 &     0.3 &     0.7\\
&  2008.860 & $-$3.2643  $\pm$  0.0005 & $-$6.9141  $\pm$  0.0006 &    $-$0.5 &    $-$0.3\\
&  2011.470 & $-$3.2686  $\pm$  0.0005 & $-$6.9232  $\pm$  0.0009 &     0.4 &    $-$0.2\\
&  2014.249 & $-$3.2717   $\pm$ 0.0002 & $-$6.9323  $\pm$  0.0005 &     0.4 &     0.6\\
&  2020.988 & $-$3.2789  $\pm$  0.0002 & $-$6.9501  $\pm$  0.0003 &    $-$0.3 &     0.3\\
&  2022.227 & $-$3.2807  $\pm$  0.0001 & $-$6.9544  $\pm$  0.0001 &     0.4 &     0.7\\
\hline
IRS 28   &   1998.410 & 10.4700  \pm  0.0030 & $-$5.7883 \pm   0.0050 &    $-$0.0 &     0.7\\
&  2000.850 & 10.4694 \pm   0.0010 & $-$5.7956 \pm   0.0024 &     0.4 &    $-$0.8\\
&  2006.200 & 10.4809 \pm   0.0007 & $-$5.8254  \pm  0.0010 &    $-$1.1 &     0.5\\
&  2008.860 & 10.4839 \pm   0.0005 & $-$5.8385 \pm   0.0006 &     0.4 &    $-$0.8\\
&  2011.470 & 10.4857 \pm   0.0005 & $-$5.8536 \pm   0.0010 &     0.4 &     0.5\\
&  2014.249 & 10.4927  \pm  0.0003 & $-$5.8693  \pm  0.0005 &    $-$0.9 &     0.7\\
&  2020.988 & 10.5018  \pm  0.0014 & $-$5.9074 \pm   0.0021 &    $-$0.5 &    $-$0.5\\
&  2022.227 & 10.5042  \pm  0.0003 & $-$5.9125 \pm   0.0006 &    $-$1.0 &     0.6\\
\hline
SiO-15   &     2000.850 & $-$12.4253 \pm   0.0023 & $-$11.0794 \pm   0.0054 &    $-$1.3 &    $-$0.2\\
&  2006.200 & $-$12.4384 \pm   0.0012 & $-$11.0753  \pm  0.0015 &     1.2 &     1.1\\
&  2008.860 & $-$12.4486  \pm  0.0019 & $-$11.0743  \pm  0.0011 &    $-$1.2 &    $-$0.1\\
&  2014.240 & $-$12.4610 \pm   0.0004 & $-$11.0702  \pm  0.0007 &     1.3 &     0.9\\
&  2020.980 & $-$12.4765 \pm  0.0007 & $-$11.0666  \pm  0.0018   &   $-$0.5  &     0.8\\
&  2022.220&  $-$12.4814 \pm  0.0005 & $-$11.0641 \pm  0.0005 &     1.3 &     1.1\\
\hline
IRS 10EE  &     1996.413 &  7.6841  \pm  0.0005 &  4.2191 \pm   0.0005 &    $-$0.4 &     0.3\\
&  1998.410 &  7.6836  \pm  0.0005 &  4.2156 \pm   0.0005 &    $-$0.5 &     0.3\\
&  2000.850 &  7.6845  \pm  0.0005 &  4.2099  \pm  0.0005 &     0.7 &    $-$0.0\\
&  2006.200 &  7.6839  \pm  0.0005 &  4.1990 \pm   0.0005 &    $-$0.7 &    $-$0.4\\
&  2008.860 &  7.6837  \pm  0.0005 &  4.1939 \pm   0.0005 &     0.7 &    $-$0.1\\
&  2011.470 &  7.6846  \pm  0.0006 &  4.1902 \pm   0.0006 &    $-$0.2 &    0.4\\
&  2014.249 &  7.6848  \pm  0.0003 &  4.1842 \pm   0.0004 &    $-$0.8 &    $-$0.3\\
&  2020.988 &  7.6850  \pm  0.0001 &  4.1702  \pm  0.0002 &     0.1 &    $-$0.5\\
&  2022.227 &  7.6857  \pm  0.0001 &  4.1680 \pm   0.0002 &    $-$0.8 &    $-$0.4\\
\hline
IRS 15NE  &     1996.413 &  1.2308 \pm   0.0005 & 11.3318 \pm   0.0005 &    $-$0.6 &    $-$0.3\\
&  1998.410 &  1.2270 \pm   0.0004 & 11.3200 \pm   0.0006 &    $-$0.6 &    $-$0.3\\
&  2000.850 &  1.2227 \pm   0.0011 & 11.3024  \pm  0.0025 &     0.6 &     0.7\\
&  2006.200 &  1.2112 \pm   0.0005 & 11.2761 \pm  0.0010 &    $-$0.1 &    $-$1.1\\
&  2008.860 &  1.2062  \pm  0.0006 & 11.2596 \pm   0.0007 &     0.6 &     0.6\\
&  2011.470 &  1.2029 \pm   0.0005 & 11.2447 \pm   0.0005 &    $-$0.6 &     0.1\\
&  2014.249 &  1.1972  \pm  0.0005 & 11.2295 \pm  0.0012 &    $-$0.3 &    $-$1.1\\
&  2020.988 &  1.1828  \pm  0.0003 & 11.1894 \pm   0.0017 &     0.6 &    $-$0.2\\
&  2022.227 &  1.1816 \pm   0.0003 & 11.1818 \pm  0.0004    & $-$0.2    & $-$1.1\\
\hline
SiO-16   &    1998.410  &$-$26.4235 \pm   0.0020  &$-$34.4517  \pm  0.0027 &     2.6 &    $-$0.3\\
&  2000.850  &$-$26.4217 \pm  0.0006  &$-$34.4413  \pm  0.0011 &    $-$3.2 &    $-$1.2\\
&  2006.200  &$-$26.4190 \pm  0.0006  &$-$34.4553 \pm  0.0013    &  2.5    &  3.5\\
&  2008.860  &$-$26.4183 \pm  0.0010  &$-$34.4473 \pm  0.0025    & $-$3.1    & $-$0.9\\
&  2011.470  &$-$26.4181 \pm   0.0008  &$-$34.4556 \pm   0.0007 &     1.7 &    $-$1.6\\
&  2014.249  &$-$26.4216 \pm  0.0005  &$-$34.4696 \pm  0.0010    &  3.0    &  3.0\\
&  2022.227 & $-$26.4205  \pm  0.0002  &$-$34.4832  \pm  0.0004 &     2.8 &     3.2\\
\hline
SiO-6    &     1998.410 & 35.2203 \pm   0.0029 & 30.6532  \pm  0.0050 &    $-$2.7 &     0.9\\
&  2000.850 & 35.2324 \pm   0.0006 & 30.6595  \pm  0.0010 &     3.6 &     0.5\\
&  2006.200 & 35.2449 \pm   0.0010 & 30.6702 \pm   0.0028 &    $-$3.4 &    $-$3.2\\
 & 2008.860 & 35.2486 \pm  0.0009 & 30.6793  \pm  0.0049 &     3.5 &     0.3\\
 & 2011.470 & 35.2658 \pm   0.0020 & 30.6878 \pm   0.0040 &    $-$1.5 &     2.1\\
&  2014.249 & 35.2725 \pm   0.0018 & 30.7022 \pm   0.0019 &    $-$3.8 &    $-$2.5\\
 & 2020.988 & 35.2838 \pm   0.0005 & 30.7059 \pm   0.0016 &     1.1 &    $-$2.5\\
&  2022.227 & 35.2945 \pm   0.0008 & 30.7144 \pm   0.0019 &    $-$3.7 &    $-$2.8\\
\hline
SiO-17   &    1998.410 &  8.0426 \pm   0.0005  & $-$27.7032 \pm  0.0014    &  1.2    &  1.3\\
&  2006.200 &  8.0623 \pm   0.0005  &$-$27.6852 \pm  0.0009    & $-$0.9    &  2.7\\
&  2008.860 &  8.0698 \pm   0.0008  &$-$27.6760 \pm  0.0004    & $-$0.7    & $-$2.0\\
&  2011.470 &  8.0719 \pm   0.0006  &$-$27.6652 \pm  0.0010    &  1.5    &  0.3\\ 
& 2014.249 &  8.0818  \pm  0.0002 & $-$27.6648 \pm  0.0004    & $-$0.4    &  2.8\\
&  2020.988 &  8.0988 \pm   0.0005 & $-$27.6453 \pm  0.0011    & $-$1.6    & $-$0.0\\
&  2022.227 &  8.1013 \pm   0.0002 & $-$27.6431 \pm  0.0005    & $-$0.6    &  2.8\\
\hline
SiO-11   &    1998.410 &  1.7380 \pm   0.0023 & 40.2675 \pm   0.0032 &    $-$2.1 &    $-$1.1\\
&  2000.850 &  1.7400 \pm   0.0005 & 40.2795 \pm   0.0011 &     1.8 &     2.6\\
&  2006.200 &  1.7462  \pm  0.0005 & 40.2914 \pm   0.0006 &    $-$0.0 &    $-$4.0\\
&  2008.860 &  1.7475  \pm  0.0003 & 40.2954  \pm  0.0008 &     1.9 &     2.4\\
&  2011.470 &  1.7590  \pm  0.0002 & 40.3034 \pm   0.0004 &    $-$2.2 &     0.3\\
&  2014.249 &  1.7610  \pm  0.0003 & 40.3119 \pm   0.0010 &    $-$0.7 &    $-$3.8\\
&  2020.988 &  1.7708 \pm   0.0003 & 40.3168 \pm   0.0005 &     2.2 &    $-$0.7\\
&  2022.227 &  1.7759 \pm   0.0003 & 40.3240 \pm   0.0005 &    $-$0.4    & $-$4.0\\
\hline
IRS 17    &   2000.850 & 13.1502  \pm  0.0026 &  5.5651  \pm  0.0025 &     1.1 &    $-$0.2\\
&  2006.200 & 13.1414  \pm  0.0013 &  5.5611 \pm   0.0021 &    $-$1.3 &    $-$0.6\\
&  2008.860 & 13.1445 \pm   0.0010 &  5.5554  \pm  0.0056 &     1.0 &    $-$0.2\\
&  2014.249 & 13.1359 \pm   0.0023 &  5.5513  \pm  0.0017 &    $-$1.3 &    $-$0.4\\
&  2020.988 & 13.1270 \pm  0.0036 &  5.5443  \pm  0.0081 &     0.1 &    $-$0.8\\
&  2022.227 & 13.1262  \pm  0.0022 &  5.5437 \pm   0.0048 &    $-$1.3 &    $-$0.5\\
\hline
SiO-12   &    2000.850 & $-$18.8236  \pm  0.0028 & 42.4687  \pm  0.0032 &     0.5 &     3.6\\
&  2006.200 & $-$18.8211 \pm   0.0017 & 42.4459  \pm  0.0050 &     2.0 &    $-$4.1\\
&  2008.860 & $-$18.8164  \pm  0.0018 & 42.4647  \pm  0.0025 &     0.7 &     3.4\\
&  2011.470  &$-$18.8091 \pm   0.0010 & 42.4721 \pm   0.0020 &    $-$2.4 &    $-$0.8\\
&  2014.249  &$-$18.8113  \pm  0.0005 & 42.4820 \pm   0.0011 &     1.2 &    $-$4.3\\
&  2020.988  &$-$18.8030  \pm  0.0007 & 42.4850  \pm  0.0027 &     2.6 &     0.4\\
&  2022.227 & $-$18.8009  \pm  0.0008 & 42.4910  \pm  0.0012    &  1.6    & $-$4.3\\
\hline
IRS 19NW  &   1998.410 & 14.5516  \pm  0.0011 & $-$18.4617\pm  0.0012    &  0.5    &  1.3\\
&  2000.850 &14.5533 \pm 0.0015 & $-$18.4683  \pm  0.0031 &     0.2 &    $-$1.8\\
&  2006.200 & 14.5606  \pm  0.0005 & $-$18.4656 \pm  0.0009    & $-$1.5    &  1.8\\
&  2008.860 &  14.5620 \pm  0.0010 & $-$18.4695  \pm  0.0013 &     0.1 &    $-$1.8\\
&  2011.470 & 14.5625 \pm   0.0004 & $-$18.4638  \pm  0.0008 &     1.1 &     0.7\\
&  2014.249 & 14.5706  \pm  0.0003& $-$18.4729 \pm  0.0007 &   $-$1.1    &  2.0\\
&  2020.988 & 14.5793 \pm  0.0003 & $-$18.4753 \pm   0.0008 &    $-$1.2 &    $-$0.5\\
&  2022.227 & 14.5815  \pm  0.0001 & $-$18.4768 \pm   0.0002 &    $-$1.3 &     1.9\\
\enddata
\tablecomments{Coordinate offsets are with respect to the Sgr~A* radio centroid at each epoch.  The relative differential offsets listed
  in the last two columns and described in Section \ref{sec:methods} have been subtracted from the coordinate offsets.}
\tablenotetext{a}{This offset is corrected for Declination:  it is $\Delta {\rm RA} \cos({\rm Dec.})$.}
\tablenotetext{b}{The first two epochs were not used for the IRS~7 astrometry but are included here for posterity.}
\end{deluxetable*}



\section{Proper Motion Measurement Methods}\label{sec:pms}

\begin{figure*}[ht!]
\begin{centering}
  \includegraphics[scale=0.5,trim= 0 44 20 0,clip=true]{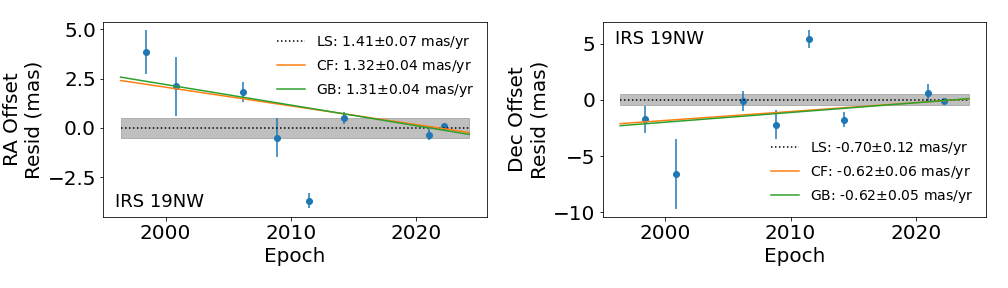}
  \includegraphics[scale=0.5,trim= 0 20 20 0,clip=false]{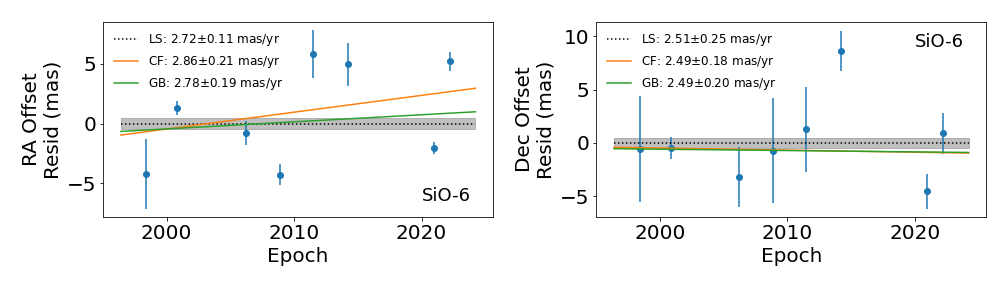}
  \caption{Example residual proper motion time series showing the three fitting methods (LS = least-squares; CF = conservative formulation; GB = good-and-bad data).
  The residual is with respect to the weighted least-squares fit, and the shaded regions indicate $\pm4$~AU.
  Top:  IRS 19NW, showing proper motion solutions for the conservative formulation and good-and-bad methods that differ from
  the canonical least-squares fit.
   Bottom: SiO-6 shows differing proper motion solutions in RA for all three methods.
  For most of the masers, the two less conventional fitting methods do not improve upon or differ significantly from 
  the least-squares method.  RA offsets are true angular offsets (i.e., they are corrected for $\cos({\rm Dec.})$).}
\label{fig:PMresids}
\end{centering}
\end{figure*}



Some of the time series in Figure \ref{fig:pms} show large single-epoch outliers from secular trends.  To address the impact of these
outliers on proper motion measurements, we 
explored three different proper motion fitting methods:  (1) variance-weighted least-squares; (2) a ``conservative formulation'' of uncertainties
\citep{sivia2006, darling2018} ; and (3) a ``good-and-bad data'' model \citep{box1968,sivia2006}.
Details on all three methods are presented here.

(1)  The weighted least-squares method  minimizes the variance-weighted difference between the model and the data.  For proper motions,
the data are simply the coordinate time series, and the model is a line (slope and intercept).  For this method and the following fitting methods,
we assign the reference epoch to be the coordinate variance-weighted mean of all observation epochs.  The coordinate variance at each epoch is the
sum in quadrature of the uncertainty in each coordinate.  For linear fitting, all epochs are relative to the reference epoch, and the intercept of the proper motion
fit is the coordinate at the reference epoch.  This approach minimizes the correlation between the slope and intercept of linear fits.
We used \texttt{lmfit}\footnote{\citet{newville2021}} for the minimization.   

We simultaneously fit the proper motions in both coordinates in order to assess correlations between fit parameters and any crosstalk between
nominally orthogonal proper motions.  There can be a minor correlation between the slope and intercept of single-coordinate fits because
one reference epoch is used for both coordinates.  
Correlation between RA and Dec.\ solutions is generally negligible.

(2) The ``conservative formulation'' does not assume Gaussian measurement uncertainties.  Instead, it
treats errorbars as lower bounds and assigns slowly decaying tails to the probability distribution, which can
reduce the impact of outlier data points \citep{sivia2006}.
In this paradigm, the probability of an error-weighted data-model residual $R_i$ for data point $i$, given a model, is 
\begin{equation}
  {\rm prob}(R_i) \propto  {1-e^{-R_i^2/2} \over R_i^2} .
\end{equation}
We maximize the sum of the logarithm of this probability (the posterior probability density) to estimate the  model parameters and their marginalized uncertainties
using \texttt{emcee}\footnote{\citet{emcee}}.  The four model parameters are generally uncorrelated.

(3) The ``good-and-bad data'' model assumes that the uncertainties in measurements are underestimated by a factor $\gamma$
with probability $\beta$ such that
\begin{equation}
  {\rm prob}(R_i) \propto \left( {\beta\over\gamma}\, e^{-R_i^2/2\gamma} + (1-\beta)\, e^{-R_i^2/2} \right)
\end{equation}
\citep{box1968,sivia2006}.
This effectively introduces two additional parameters to the proper motion fits, for a total of six (a slope and intercept for each coordinate, $\gamma$, and $\beta$).
It is worth noting that this method does not identify individual ``bad'' datapoints.  

As with the conservative formulation, we maximize the sum of the logarithm of the probability using an MCMC process.   We place bounds
(uniform priors) on the two good-and-bad parameters:  $0 \leq \beta \leq 1$ and $1 \leq \gamma \leq 10$.  $\beta$ and $\gamma$ tend to be anti-correlated (a lower probability of bad data drives larger
uncertainties) and are often correlated with the linear fit parameters.  Values for $\beta$ are typically 0.3--0.8 and for $\gamma$ are 2--4, which represent a
high probability that the data have variances a factor of several too small.   Nevertheless, this method does not typically produce significantly different proper motion measurements than the simple weighted least-squares method.
There are two masers with large outliers where this method does favor large uncertainty scaling $\gamma\sim7$--8 and proper motion solutions
that differ significantly from the other two methods, namely SiO-11 and IRS~19NW.
SiO-6 has $\gamma\sim 5$ and shows a bimodal likelihood distribution in RA slope and offset in both alternative methods,
but all three slopes are consistent within their error budgets.  
Figure \ref{fig:PMresids} shows the proper motion fit residuals for IRS~19NW and SiO-6.  
Note that we did not assign separate $\beta$ and $\gamma$ parameters to each coordinate time series.

The majority of maser proper motions obtained from these three fitting methods are indistinguishable given their uncertainties,
so we report the least-squares fits in Table \ref{tab:results} and Figure \ref{fig:pms}.
Because the conservative formulation and good-and-bad data methods allow for larger uncertainties in the data,
they would normally be expected to produce larger uncertainties in parameter estimates than variance-weighted least squares fits.
However, contrary to what is expected for uniform Gaussian random errors, the proper motion uncertainties
are typically, but not exclusively, larger for the least-squares method compared to the other two methods.
When $\chi^2_\nu \sim 1$ (see Table \ref{tab:results}), uncertainties in the alternative methods are larger than
for the least-squares estimates, following the canonical expectation.

\section{Expanded Astrometric Uncertainties}\label{sec:errors}

 To assess the large astrometric departures from secular trends seen in Figure \ref{fig:pms}
  we quantify a time-independent additional uncertainty that could be added in quadrature to the
  measurement uncertainties quoted in Table \ref{tab:time_series}.  To do this, we adopt a linear fit model
  that includes an additional uncertainty parameter for each coordinate.  Starting with the variance-weighted
  least-squares solutions, we perform an initial (log) likelihood maximization fit followed by
  an MCMC exploration of the parameter space using \texttt{emcee} \citep{emcee}
to estimate uncertainties in the new fit parameters.  This process
  effectively finds the additional constant uncertainty that would be included in the time series in order to
maximize the likelihood of a linear fit.  

For three (20\%) of the masers, the expanded uncertainties are negligible (less than 0.1~mas) compared to the measurement
uncertainties in both coordinates.
For six (40\%) of the masers, one of the two coordinates favors a significant increase in astrometric uncertainty.  These range from
0.3 to 2.1~mas, and favor the RA direction over Dec.\ (4 vs.\ 2).  
For the remaining six (40\%), both coordinate time series have significant uncertainty increases, 
spanning 1.7--8.5 mas, and are always larger for Dec.\ compared to RA.  
In the latter two groups, however, the proper motions obtained from the MCMC process that allows for expanded uniform (time-independent) uncertainties are formally consistent with the least-squares proper motions that rely on the original measurement uncertainties.  This treatment does not obtain different proper
motions from the basic least-squares method.

\begin{figure*}[ht!]
\begin{centering}
  \includegraphics[scale=0.5,trim= 0 44 20 0,clip=true]{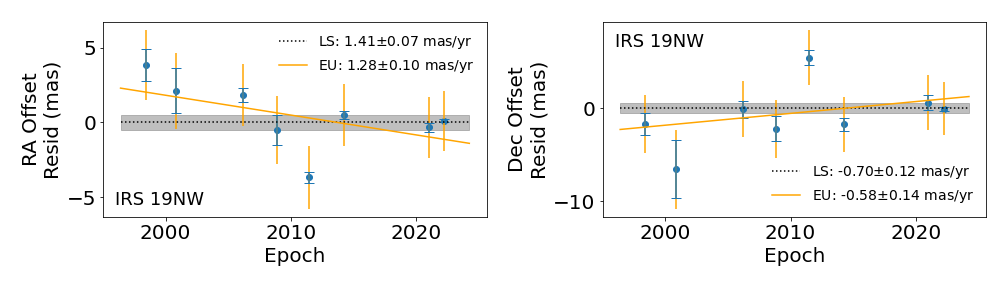}
  \includegraphics[scale=0.5,trim= 0 20 20 0,clip=false]{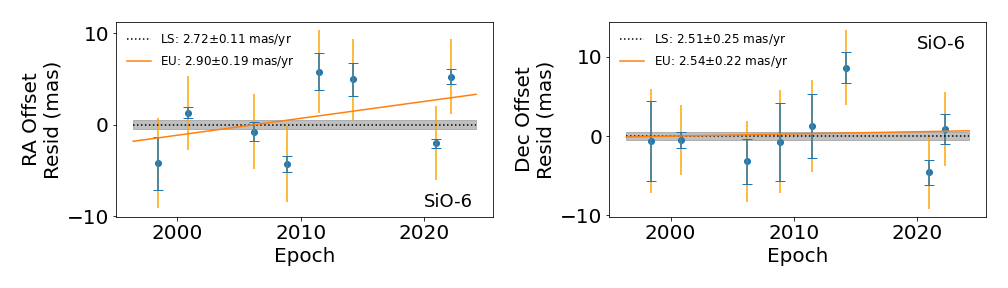}
  \caption{Example residual proper motion time series showing the result of 
    expanded uncertainties (EU) that are added in quadrature to the measurement uncertainties (see Appendix \ref{sec:errors}).  
  The residual is with respect to the weighted least-squares (LS) fit, and the shaded regions indicate $\pm4$~AU.
  Top:  IRS 19NW shows proper motion solutions that differ from the canonical least-squares fits but are formally consistent
  given the uncertainties.  The additional uncertainties are 2.0 mas in RA and 2.8 mas in Dec.  
  Bottom: SiO-6 shows a differing proper motion solution in RA but not in Dec, which have expanded uncertainties
  of 4.0 and 4.3 mas, respectively.  RA offsets are true angular offsets (i.e., they are corrected for $\cos({\rm Dec.})$).}
\label{fig:PMresidsEU}
\end{centering}
\end{figure*}

Figure \ref{fig:PMresidsEU} shows example fits for the same masers depicted in Figure \ref{fig:PMresids}, 
  IRS~19NW and SiO-6.  The proper motions are not formally different from the least-squares values, but it is clear that the expanded astrometric uncertainties are much larger than the expected size of the maser-emitting region in these stars, suggesting that the departures from the linear trends are not simply due to the changing structure of the
  maser-emitting regions.

\end{document}